\newif\iffullversion
\fullversiontrue
\iffullversion
\documentclass[acmsmall,nonacm]{acmart}
\else
\documentclass[acmsmall]{acmart}
\fi

\usepackage[utf8]{inputenc}
\usepackage[T1]{fontenc}

\usepackage{prooftree}
\usepackage{amstext}
\usepackage{xcolor} %

\usepackage[normalem]{ulem}
\usepackage{textcomp}
\IfFileExists{luximono.sty}{\usepackage[scaled=0.9]{luximono}}{\usepackage[scaled=0.9]{beramono}}
\usepackage{afterpage}
\usepackage{graphicx}
\usepackage{latexsym}
\usepackage{url}
\usepackage{amsmath}
\usepackage{listings}
\definecolor{darkgreen}{RGB}{34,149,34}

\newcommand{\figref}[1]{\figurename~\ref{fig:#1}}
\newcommand{\secref}[1]{Sec.~\ref{sec:#1}}
\newcommand{\appref}[1]{App.~\ref{app:#1}}

\newcommand{\code}[1]{\text{\lstinline[basicstyle={\sffamily}]!#1!}}

\makeatletter

\newcommand{\multOpSym}{\ensuremath{\odot}}
\newcommand{\mult}{\ensuremath{\@ifnextchar\bgroup{\multT}{\multOpSym}}}
\newcommand{\multT}[2]{\ensuremath{\code{{#2}}}}

\def\old{\operator{\code{old}}}
\def\oldl#1{\operator{\ensuremath{\code{old}_{#1}}}}

\newcommand{\fullversionref}[1]{%
\iffullversion%
\appref{#1}%
\else
\citet{arxiv}%
\fi%
}

\newskip \point

\setbox136=\hbox{\leavevmode\raise0\point\hbox{${\lfloor}\kern-3\point{\lfloor}$}}
\setbox137=\hbox{\raise0\point\hbox{${ \rfloor}\kern-3\point{ \rfloor}$}}

\makeatother

\newcommand{\myparagraph}[1]{\textit{#1}}

\newcommand{\eg}{{{e.g.\@}}} 
\newcommand{\ie}{{{i.e.\@}}} 

\newcommand{\wrt}{w.r.t.\@}

\makeatletter
\def\operator#1{\@ifnextchar\bgroup {\operatorarg{\ensuremath{#1}}}{\ensuremath{#1}}}
\def\operatorarg#1#2{{#1}{\ensuremath{(#2)}}}
\def\spoperator#1{\@ifnextchar\bgroup{\spoperatorarg{\ensuremath{#1}}}{\ensuremath{#1}}}
\def\spoperatorarg#1#2{\ensuremath{#1~#2}}

\def\method#1{\@ifnextchar\bgroup{\methodarg{\code{#1}}}{\code{#1}\xspace}}
\def\methodarg#1#2{{#1}\code{(#2)}\xspace}

\def\FV{\operator{\textit{FV}}}
\def\mods{\operator{\textit{mods}}}

\newcommand{\preL}{\textit{pre}}
\newcommand{\lastL}{\textit{last}}
\newcommand{\callL}{\textit{call}}

\makeatother

\newcommand\stackCloserHelp[2]{{%
    \setbox0\hbox{\ensuremath{#1}}%
    \rlap{\hbox to \wd0{\hss\raisebox{-.7\height}{#2}\hss}}\box0
}}

\makeatletter

\newcommand\noop{\ensuremath{\mathtt{skip}}}

\newcommand\assign[2]{\ensuremath{#1{:}{=}#2}}

\newcommand\new[2]{\@ifnextchar\bgroup{\newthree{#1}{#2}}{\newthree{#1}{#2}{}}}

\newcommand\newthree[3]{\ensuremath{\assign{#1}{\mathtt{~spawn~}#2(#3)}}}

\newcommand{\triple}[3]{\@ifnextchar\bgroup{\tripleFour{#1}{#2}{#3}}{\tripleFour{\env}{#1}{#2}{#3}}}
\newcommand{\tripleFour}[4]{\ensuremath{{#1}\vdash \{#2\}~#3~\{#4\}}}
\newcommand{\env}{\Lambda}

\usepackage{relsize}

\makeatletter

\newcommand{\judgementdef}[3]{\ensuremath{#2#1 #3}}
\newcommand{\judgement}[1]{\@ifnextchar\bgroup{\judgementArg{#1}}{\judgementdef{#1}{}{}}}
\newcommand{\judgementArg}[2]{\@ifnextchar\bgroup{\judgementArgs{#1}{#2}}{\judgementdef{#1}{}{#2}}}
\newcommand{\judgementArgs}[3]{\@ifnextchar\bgroup{#2,\judgementArgs{#1}{#3}}{\judgementdef{#1}{#2}{#3}}}

\newcommand{\entails}{\judgement{\models}}

\newcommand{\OKsym}{\ensuremath{\vdash_{\textit{OK}}}}
\newcommand{\OK}{\@ifnextchar\bgroup{\OKone}{\ensuremath{\OKsym}}}
\newcommand{\OKone}[1]{\@ifnextchar\bgroup{\OKmore{#1}}{\ensuremath{#1 \OKsym}}}
\newcommand{\OKmore}[2]{\@ifnextchar\bgroup{\OKthree{#1,#2}}{\OKone{#1,#2}}}
\newcommand{\OKthree}[3]{\ensuremath{#1,#2\OKsym #3}}

\newcommand{\futureEntailsSym}{\textit{futureEntails}}
\newcommand{\futureEntails}{\@ifnextchar\bgroup{\futureEntailsArg}{\futureEntailsSym}}
\newcommand{\futureEntailsArg}[3]{\@ifnextchar\bgroup{\futureEntailsArg{#1,#2}{#3}}{\ensuremath{(#1){.}\futureEntailsSym(#2,#3)}}}

\newcommand{\futureCombinesSym}{\textit{futureCombines}}
\newcommand{\futureCombines}{\@ifnextchar\bgroup{\futureCombinesArg}{\futureCombinesSym}}
\newcommand{\futureCombinesArg}[4]{\@ifnextchar\bgroup{\futureCombinesArg{#1,#2}{#3}{#4}}{\ensuremath{(#1){.}\futureCombinesSym(#2,#3,#4)}}}

\newcommand{\freeze}{\@ifnextchar\bgroup{\freezeone}{\ensuremath{\textit{freeze}}}}
\newcommand{\freezeone}[1]{\@ifnextchar\bgroup{\freezemore{#1}}{\ensuremath{\textit{freeze}(#1)}}}
\newcommand{\freezemore}[2]{\freeze{#1,#2}}

\def\call#1#2{\@ifnextchar\bgroup{\calldef{#1}{#2}}{\calldef{#1}{#2}{}}} 
\def\calldef#1#2#3{\@ifnextchar\bgroup{\callmergeargs{#1}{#2}{#3}}{\deref{#1}{{#2}({#3})}}}
\def\callmergeargs#1#2#3#4{\calldef{#1}{#2}{#3,#4}}
\def\deref#1#2{\ensuremath{#1{.}#2}}

\newcommand{\response}[2]{\@ifnextchar\bgroup{\responsethree{#1}{#2}}{\call{#1}{#2}}}
\newcommand{\responsethree}[3]{\@ifnextchar\bgroup{\responsewhere{#1}{#2}{#3}}{\responsewhere{#1}{#2}{}{#3}}} 
\newcommand{\responsewhere}[4]{\call{#1}{#2}{#3}\;\where\;{#4}}
\newcommand{\where}{\textit{where}}

\usepackage{xspace}

\ifx\symlasy\undefined   \DeclareSymbolFont{lasy}{U}{lasy}{m}{n}
  \SetSymbolFont{lasy}{bold}{U}{lasy}{b}{n} \else
\fi
\DeclareMathSymbol\safeleadsto {\mathrel}{lasy}{"3B}

\newskip \point

\setbox136=\hbox{\leavevmode\raise0\point\hbox{${\lfloor}\kern-3\point{\lfloor}$}}
\setbox137=\hbox{\raise0\point\hbox{${ \rfloor}\kern-3\point{ \rfloor}$}}

\setbox138=\hbox{\raise1.5\point\hbox{${ \safeleadsto}\kern-10\point$}\raise-1.5\point\hbox{${ \safeleadsto}$}}

\def \premisespacing{\quad\;}

\newcommand{\rulename}[1]{\ensuremath{(\textit{#1})}}

\def \RulePremisesNewlineMore[#1]#2.#3#4{\@ifnextchar\bgroup{\RulePremisesNewlineMore[#1]{#2}.{#3\premisespacing#4}}{\@ifnextchar.{\RulePremisesNewline[#1]{{\begin{array}{c}#2\\#3\premisespacing#4\end{array}}}}{\RuleMultiPremise[#1]{{\begin{array}{c}#2\\#3\end{array}}}{#4}}}}

\def \RulePremisesNewline[#1]#2.#3{\@ifnextchar\bgroup{\RulePremisesNewlineMore[#1]{#2}.{#3}}{\@ifnextchar.{\RulePremisesNewline[#1]{{\begin{array}{c}#2\\#3\end{array}}}}{\RuleMultiPremise[#1]{#2}{#3}}}}

\def \RuleMultiPremise[#1]#2#3{\@ifnextchar\bgroup{\RuleMultiPremise[#1]{#2\premisespacing#3}}{\@ifnextchar.{\RulePremisesNewline[#1]{#2\premisespacing#3}}{\prooftree #2\justifies#3 \using{#1}\endprooftree}}}

\def \RuleWithName[#1]#2{\@ifnextchar\bgroup {\RuleMultiPremise[#1]{#2}}{\@ifnextchar.{\RulePremisesNewline[#1]{#2}}{\prooftree \justifies #2 \using{#1} \endprooftree}}}

\def \RuleWithInfo[#1]{\@ifnextchar[{\RuleWithNameAndCondition[#1]}{\RuleWithName[(\textit{#1})]}}

\def \RuleWithNameAndCondition[#1][#2]{\RuleWithName[\rulename{#1}^{#2}]}

\def \Inf{\proofrulebaseline=2ex \abovedisplayskip12\point\belowdisplayskip12\point \abovedisplayshortskip8\point\belowdisplayshortskip8\point \@ifnextchar[{\RuleWithInfo}{\RuleWithName[ ]}}

\makeatother

\usepackage{pstricks}
\def \longharpoon#1{\psset{unit=1\point,linewidth=0.35\point}%
\psline{cc-cc}(0,1.5)(.9#1,1.5)%
\hspace*{.9#1}\pscurve{cc-cc}(0,1.5)(-1.1,2)(-2.25,3.7)\hspace*{.1#1}}

\def \Vec#1{\setbox155=\hbox{$#1$}%
\leavevmode\copy155\kern-.95\wd155
\raise\ht155\hbox{\longharpoon{\wd155}}}

\newcommand\eeval[5]{\ensuremath{ \llbracket #1 \rrbracket_{(#2, #3, #4, #5)}}}
\newcommand{\wand}{\ensuremath{\mathbin{-\mkern-7mu-\mkern-8mu*}}}
\newcommand\twovyper{\textsc{2Vyper}\xspace}
\newcommand\cassert[1]{\ensuremath{\code{assert}~#1}}

\newcommand\cseq[2]{\ensuremath{#1; #2}}
\newcommand\cassign[2]{\ensuremath{#1 := #2}}
\newcommand\ccall[5]{\ensuremath{#1 := #2.\code{#3}(#4, \code{value}=#5)}}
\newcommand\ccalltl[5]{\ensuremath{
\begin{array}{c}
#1 := #2.\code{#3}\\
(#4, \code{value}=#5)
\end{array}
}}
\newcommand\mcup{\ensuremath{\cup^{\#}}}
\newcommand\aemp{\ensuremath{\code{emp}}}
\newcommand\tru{\ensuremath{\code{true}}}
\newcommand\fal{\ensuremath{\code{false}}}
\newcommand\stable[2]{\ensuremath{\code{stable}(#1,#2)}}
\newcommand\self{\ensuremath{\code{self}}}
\newcommand\msg{\ensuremath{\code{msg}}}
\newcommand\block{\ensuremath{\code{block}}}
\newcommand\result{\ensuremath{\code{result}}}
\newcommand\CInv{\ensuremath{\code{TSC}}}
\newcommand\GInv{\ensuremath{\code{ITSC}}}
\newcommand\CLC{\ensuremath{\code{SC}}}
\newcommand\ContractState[1]{\ensuremath{\code{CS}_{#1}}}
\newcommand\ensures[1]{\ensuremath{\code{ensures}~#1}}
\newcommand\performs[1]{\ensuremath{\code{performs}~#1}}
\newcommand\multiset[1]{\ensuremath{\{ #1 \}^{\#}}}

\newcommand\aperformed[1]{\ensuremath{\code{perf}(#1)}}
\newcommand\fulfils[3]{\ensuremath{#1, #2 \models #3}}
\newcommand\CGPost{\ensuremath{\code{FC}}}
\newcommand\CCallerPrivate[2]{\ensuremath{\code{PC}(#1, #2)}}
\newcommand\derivedDestroyed[1]{\ensuremath{\code{derDestroyed}(#1)}}
\newcommand\derivedCreated[1]{\ensuremath{\code{derCreated}(#1)}}
\newcommand\derivedPerformed[1]{\ensuremath{\code{derPerformed}(#1)}}
\newcommand\creallocate[4]{\ensuremath{\code{transfer}_{#1}(#2, #3, #4)}}
\newcommand\ereallocate[4]{\ensuremath{\code{transfer}_{#1}(#2, #3, #4)}}
\newcommand\coffer[7]{\ensuremath{\code{offer}_{#1 \leftrightarrow #2}(#3, #4, #5, #6, #7)}}
\newcommand\coffertl[7]{\ensuremath{
\begin{array}{c}
\code{offer}_{#1 \leftrightarrow #2}\\
(#3, #4, #5, #6, #7)
\end{array}
}}
\newcommand\eoffer[7]{\ensuremath{\code{offer}_{#1 \leftrightarrow #2}(#3, #4, #5, #6, #7)}}
\newcommand\crevoke[7]{\ensuremath{\code{revoke}_{#1 \leftrightarrow #2}(#3, #4, #5, #6, #7)}}
\newcommand\crevoketl[7]{\ensuremath{
\begin{array}{c}
\code{revoke}_{#1 \leftrightarrow #2} \\
(#3, #4, #5, #6, #7)
\end{array}
}}

\newcommand\cexchange[6]{\ensuremath{\code{exchange}_{#1 \leftrightarrow #2}(#3, #4, #5, #6)}}
\newcommand\cexchangetl[6]{\ensuremath{
\begin{array}{c}
\code{exchange}_{#1 \leftrightarrow #2} \\
(#3, #4, #5, #6)
\end{array}
}}
\newcommand\eexchange[7]{\ensuremath{\code{exchange}_{#1 \leftrightarrow #2}(#3, #4, #5, #6, #7)}}
\newcommand\ctrust[2]{\ensuremath{\code{trust}(#1, #2)}}
\newcommand\etrust[4]{\ensuremath{\code{trust}_{#4}(#1, #2, #3)}}
\newcommand\ccreate[4]{\ensuremath{\code{create}_{#1}(#2, #3, #4)}}
\newcommand\ecreate[3]{\ensuremath{\code{create}_{#1}(#2, #3)}}
\newcommand\cdestroy[3]{\ensuremath{\code{destroy}_{#1}(#2, #3)}}
\newcommand\edestroy[3]{\ensuremath{\code{destroy}_{#1}(#2, #3)}}

\newcommand\mallocated[1]{\ensuremath{\code{balances}_{#1}}}
\newcommand\aalloc[3]{\ensuremath{\code{owns}_{#1}(#2, #3)}}

\newcommand\atrusts[3]{\ensuremath{\code{trusts}(#1, #2, #3)}}
\newcommand\mtrusted{\ensuremath{\code{trusted}}}
\newcommand\emptymultiset{\ensuremath{\emptyset^{\#}}}

\newcommand\aoffers[7]{\ensuremath{\code{offers}_{#1 \leftrightarrow #2}(#3, #4, #5, #6, #7)}}
\newcommand\moffered[2]{\ensuremath{\code{offered}_{#1 \leftrightarrow #2}}}
\newcommand\htriple[4]{\ensuremath{#1 \vdash \left \{ {#2} \right \} {#3} \left \{ {#4} \right \} }}

\newcommand\secondary[1]{\ensuremath{\mathit{secondary}(#1)}}

\newcommand\emptyr{\ensuremath{\mathcal{R}_{\mathit{empty}}}}
\newcommand\wei{\ensuremath{\code{wei}}}
\newcommand\rcreator[1]{\ensuremath{\code{creator}(#1)}}
\newcommand\rcreators{\ensuremath{\code{CREATORS}}}
\newcommand\sdefault[1]{\ensuremath{\code{default}(#1)}}

\AtBeginDocument{%
  \providecommand\BibTeX{{%
    \normalfont B\kern-0.5em{\scshape i\kern-0.25em b}\kern-0.8em\TeX}}}

\setcopyright{rightsretained}
\acmJournal{PACMPL}
\acmYear{2021} 
\acmNumber{OOPSLA} 

\begin{document}
\title{Rich Specifications for Ethereum Smart Contract Verification}

\author{Christian Br\"am}
\email{c.braem@gmx.ch}
\affiliation{
  \department{Department of Computer Science}
  \institution{ETH Zurich}
  \country{Switzerland}
}
\author{Marco Eilers}
\orcid{1234-5678-9012}
\email{marco.eilers@inf.ethz.ch}
\affiliation{
  \department{Department of Computer Science}
  \institution{ETH Zurich}
  \country{Switzerland}
}
\author{Peter M\"uller}
\email{peter.mueller@inf.ethz.ch}
\affiliation{
  \department{Department of Computer Science}
  \institution{ETH Zurich}
  \country{Switzerland}
}
\author{Robin Sierra}
\email{robin.sierra@outlook.com}
\affiliation{
  \department{Department of Computer Science}
  \institution{ETH Zurich}
  \country{Switzerland}
}
\author{Alexander J. Summers}
\email{alex.summers@ubc.ca}
\affiliation{
  \department{Department of Computer Science}
  \institution{University of British Columbia}
  \country{Canada}
}

\begin{abstract}
Smart contracts are programs that execute in blockchains such as Ethereum to manipulate digital assets. Since bugs in smart contracts may lead to substantial financial losses, there is considerable interest in formally proving their correctness. However, the specification and verification of smart contracts faces challenges that rarely arise in other application domains. Smart contracts frequently interact with unverified, potentially adversarial outside code, which substantially weakens the assumptions that formal analyses can (soundly) make. Moreover, the core functionality of smart contracts is to manipulate and transfer resources; describing this functionality concisely requires dedicated specification support. Current reasoning techniques do not fully address these challenges, being restricted in their scope or expressiveness (in particular, in the presence of re-entrant calls), and offering limited means of expressing the resource transfers a contract performs.

In this paper, we present a novel specification methodology tailored to the domain of smart contracts. Our specifications and associated reasoning technique are the first to enable: (1)~sound and precise reasoning in the presence of unverified code and arbitrary re-entrancy, (2)~modular reasoning about collaborating smart contracts, and (3)~domain-specific specifications for resources and resource transfers, expressing a contract's behaviour in intuitive and concise ways and excluding typical errors by default.
We have implemented our approach in \twovyper, an SMT-based automated verification tool for Ethereum smart contracts written in Vyper, and demonstrated its effectiveness for verifying strong correctness guarantees for real-world contracts.

\end{abstract}



\maketitle              %

\section{Introduction}

Smart contracts are programs that execute in blockchains such as Ethereum, and allow the execution of resource transactions between different parties without the need for a trusted third party. Smart contracts tend to be comparatively short but non-trivial programs, and since any bugs can and do frequently lead to the loss of potentially-large amounts of money~\cite{dao}, using formal methods to ensure their correctness is both practically viable and highly desirable.

Compared to other programs, smart contracts bring reasoning challenges that are insufficiently supported by classical verification techniques:
First, smart contracts usually call other smart contracts, for instance, to perform transactions. These other contracts are typically developed by unknown parties and cannot be assumed to be verified; they might even exhibit adversarial behaviour to gain a financial advantage. As a result, standard modular reasoning techniques such as separation logic~\cite{sl}, which reason about calls under the assumption that \emph{all} code is verified,
do not apply in this setting. This problem is exacerbated by the presence of \emph{re-entrant} calls, \ie{}
callbacks from functions called by the contract itself. A sound reasoning technique must account for \emph{all} behaviours a call to an unverified contract could possibly exhibit, which requires novel ways of specifying the calling contract.
Second, even in this adversarial setting, multiple smart contracts may form a distributed application designed to collaboratively maintain invariants across the contracts' states. Such collaborations are often decoupled using interface abstractions, which must be equipped with sufficient specification to soundly guarantee the preservation of these invariants.
Third, typical smart contracts are primarily concerned with modelling and executing \emph{resource transactions} of different kinds, ranging from token contracts to escrow implementations, ICOs, and complex decentralised finance (DeFi) applications. However, even though notions such as resource ownership and agreed exchanges are central to programmer intentions, resources themselves are often implicit in smart contract implementations. This discrepancy between high-level intentions and low-level implementations easily leads to subtle and potentially-costly mistakes.

Existing (automated) verifiers for smart contracts do not fully address these challenges. Some verifiers~\cite{DBLP:conf/icse/TikhomirovVITMA18,DBLP:conf/icse/FeistGG19,DBLP:conf/ccs/TsankovDDGBV18,DBLP:conf/iccsp/LaiL20} prove specific properties, \eg{} the absence of overflows or re-entrancy bugs, but cannot be used to prove full functional correctness of a contract.
Other verifiers~\cite{DBLP:journals/corr/abs-1907-04262,verx,DBLP:conf/ndss/KalraGDS18,DBLP:conf/csfw/HildenbrandtSRZ18} aim to prove arbitrary, user-defined properties using variations of established specification and verification techniques.
However, because these techniques are not sufficiently adapted to the setting of smart contracts, they are either not generally applicable or very imprecise in the presence of arbitrary re-entrancy. In general, no existing technique allows reasoning modularly about compositions of multiple smart contracts while preserving interface abstractions. Furthermore, existing techniques offer limited support for specification and reasoning in terms of high-level notions of custom resources such as tokens.

In this paper, we propose a novel specification and verification methodology for the sound, unbounded verification of general (safety) properties of Ethereum smart contracts. We offer specification constructs tailored to the domain of smart contracts, enabling users to prove strong functional correctness properties of arbitrary smart contracts, with specifications that capture their intended resource manipulations explicitly.
We make the following four main contributions:

\emph{(1)~Reasoning in the presence of unverified code.}
To the best of our knowledge, we present the first smart-contract verification technique that is sound and precise in the presence of calls to unverified contracts with arbitrary re-entrancy. Our technique can prove that properties cannot be invalidated by calls to unverified code, including vital properties such as access control to resources.

\emph{(2)~Modular reasoning about collaborating smart contracts.}
We explain the challenges of verifying collaborating contracts; our specification methodology can capture all required information at the interface level, providing the first \emph{modular} verification technique for collaborating smart contracts.

\emph{(3)~Native resource-oriented specifications.}
We introduce novel specification features for directly capturing resource ownership, transfer, offers to exchange, and loaning, enabling direct and concise specification of programmer intentions.
Ubiquitous properties of resources such as ownership, access control, and non-duplicability are baked into our system, avoiding potentially repetitive and error-prone boilerplate specifications; violations of these properties are found by default.

\emph{(4)~Implementation and evaluation.}
We implemented our approach in \twovyper, an automated, SMT-based verification tool for the Vyper language~\cite{vyper} for Ethereum smart contracts. It supports the entire current Vyper language and allows specifying contracts and interfaces in the form of readable, source-level code annotations.
Our evaluation shows that \twovyper enables automated verification of strong correctness properties of (collaborating) real-world contracts with reasonable performance and annotation overhead.
In particular, we demonstrate that \twovyper can verify contracts that use re-entrancy patterns unsupported by other verification tools, and that it enables modular verification of collaborating smart contracts used in practice. 
\iffullversion \else
Our implementation and evaluation are available as an artifact~\cite{artifact}. \fi

\subsubsection*{Outline.}
The paper is structured as follows: We introduce Ethereum smart contracts in Sec.~\ref{sec:prelims}. In Sec.~\ref{sec:unverified}, we informally introduce the specification constructs we use to reason about contracts containing re-entrant calls; subsequently, we show how they can be used to reason about collaborating contracts in Sec.~\ref{sec:ici}. We introduce our resource-based specification approach in Sec.~\ref{sec:resources}, and present our verification technique in the form of a Hoare logic in Sec.~\ref{sec:logic}. We describe our implementation in \twovyper and evaluate it in Sec.~\ref{sec:evaluation}. We discuss related work in Sec.~\ref{sec:related} and conclude in Sec.~\ref{sec:conclusion}.

\begin{figure}[t]
\begin{center}
\begin{minipage}[t]{0.75\textwidth}
\begin{lstlisting}
beneficiary: address
highestBid: int256
highestBidder: address
ended: bool
pendingReturns: map(address, int256)
lock: bool

def bid():
  assert not self.lock and msg.value > self.highestBid and not self.ended
  self.pendingReturns[self.highestBidder] += self.highestBid
  self.highestBidder = msg.sender
  self.highestBid = msg.value

def withdraw():
  assert not self.lock
  toSend = self.pendingReturns[msg.sender]
  self.pendingReturns[msg.sender] = 0
  self.lock = True
  send(msg.sender, value=toSend)
  self.lock = False

def end():
  assert not self.lock and not self.ended and msg.sender == self.beneficiary
  self.ended = True
  self.lock = True
  send(self.beneficiary, value=self.highestBid)
  self.lock = False
  self.highestBid = 0
\end{lstlisting}
\end{minipage}
\end{center}
\caption{Simplified auction contract written in Vyper. \code{assert} commands revert the current transaction, whereas \code{send} commands send Ether to another contract. Since the contract does not have an explicit constructor function, all fields are initialized with default values.} \label{fig:exampleauction}
\end{figure}

\section{Ethereum Smart Contracts}\label{sec:prelims}

Ethereum smart contracts are programs usually written in a high-level language, most-commonly Solidity~\cite{solidity} or the newer Vyper~\cite{vyper} language, and then compiled to bytecode for execution in the Ethereum Virtual Machine (EVM)~\cite{wood2014ethereum}.
Fig.~\ref{fig:exampleauction} shows an example of a Vyper smart contract implementing an auction. Note that in this example and throughout the paper, we use a simplified presentation of Vyper and Ethereum contracts and omit details that are irrelevant to our approach (\eg{} that Ether can be transferred only by calling functions marked as payable, or that Vyper functions revert when encountering under- or overflows\footnote{Our tool nonetheless allows one to verify that a function does not revert due to under- or overflows.}). We also ignore the fact that contract execution consumes \emph{gas}, \ie{} a fixed cost associated with every executed instruction, which is not relevant for proving safety properties, the focus of this paper.

A contract can declare \emph{fields} that form its persistent state. In our example, the contract stores the beneficiary of the auction, the current highest bid and bidder, and the amounts of \emph{wei}, a sub-unit of Ethereum's built-in currency \emph{Ether}, it owes to bidders who have been outbid.
In addition to explicitly declared fields, every contract has a built-in \code{balance} field that tracks the amount of Ether currently held by the contract.
Unlike ordinary fields, which can be written to directly by the contract (but, crucially, \emph{not} by other contracts), the \code{balance} cannot be written to directly. Ether (and wei) is the only resource with native language support; programmers can, however, implement smart contracts that provide custom resources (often called \emph{tokens}), illustrated later in this section.

Contracts define a set of functions and a special constructor function  called \code{__init__} that is executed when the contract is set up.
Smart contracts are executed as \emph{transactions}: a caller outside the blockchain can request to invoke a contract's function, and miners can then decide to execute this function as part of the next block. If this happens, the function is executed, and may in turn call functions of the same or other contracts as part of the same transaction\footnote{Throughout this paper, when we say that a contract interacts with other contracts, we mean other contract \emph{instances}, \ie{} contracts deployed at other addresses that may contain the same or (usually) different code from our contract.}. (Note that throughout this paper, we inline internal calls to private functions for simplicity.)
External calls typically occur via \emph{interfaces} that list (a subset of) the available functions of a contract. %
Importantly, there is no observable concurrency while all transitively-called functions are executed.

The intended workflow of the auction contract is that clients call the \code{bid} function and transfer along a larger amount of Ether than the current highest bid. If another client bids a higher value later, the contract updates \code{pendingReturns} to remember that no-longer-highest bidders can get their Ether back. Such a bidder can call \code{withdraw} to have the Ether transferred back to them.
Contracts can transfer Ether to other contracts via \code{send} commands (\eg{} in function \code{end}), where the parameter \code{value} specifies the transferred amount, or by calling a function (like the \code{bid} function) on another contract and implicitly passing along some amount of Ether. Internally, these are the same: executing a \code{send} command is implemented by calling a default function on the recipient.

\begin{figure}[t]
\begin{center}
\begin{minipage}[t]{0.6\textwidth}
\begin{lstlisting}
minter: address
balances: map(address, int256)

def transfer(from: address, to: address, amount: uint256):
  assert self.balances[from] >= amount and msg.sender == from
  newAmount: int256 = self.balances[from] - amount
  self.balances[to] += amount
  self.balances[from] = newAmount
  to.notify(from, self, amount)

def mint(to: address, amount: uint256):
  assert msg.sender == self.minter
  self.balances[to] += amount
\end{lstlisting}
\end{minipage}
\end{center}
\caption{Simplified token contract implemented in Vyper. The minter can create new tokens by calling \code{mint}; other users call \code{transfer} to give their own tokens to another user. \code{assert} statements ensure that the transaction reverts if a user tries to spend tokens they do not own.} \label{fig:exampletoken}
\end{figure}

Ethereum transactions can \emph{revert}, meaning they abort and all state changes they made are reset,
for several reasons. Smart contracts commonly use $\code{assert}$ commands to revert a transaction if the asserted condition is false. This is intentionally-possible behaviour used to enforce that \eg{} arguments supplied to a call are valid and that the call is allowed given the current state of the called contract. For example, a call to the \code{end} function will revert if the auction is already over, and \code{bid} reverts if the new bid is not higher than the current highest bid.
This contract also reverts if called while the \code{lock} field is set, a pattern commonly used to explicitly prevent a contract from being called in unexpected situations (often to prevent re-entrancy vulnerabilities, discussed below).

In addition to the contract's fields and explicitly declared arguments, a contract can always access the implicit arguments \code{msg} and \code{block}, which contain information about the current call and the block the current transaction is a part of. For example, \code{msg} has the particularly important field \code{msg.sender}, which contains the address of the caller of the current function. Function \code{end} uses this variable to ensure that only the beneficiary of  the auction can end the auction, whereas the \code{bid} function uses \code{msg.value} to obtain the amount of Ether sent with the call.

\myparagraph{Custom resources.}
While the auction contract works directly with the built-in Ether currency, many real contracts implement or work with \emph{tokens}~\cite{ERC20}, \ie{} custom currencies tracked via ad-hoc implementations in smart contract fields. Fig.~\ref{fig:exampletoken} shows a very simple version of a token contract. Its state consists of a map that represents the balances that each other contract holds for this token. Contracts can call \code{transfer} to transfer tokens from one contract to another, which simply corresponds to updating the map.
This contract enforces important properties common to resources in general:
Each client holding a balance should \emph{be able to transfer only tokens that it owns}. This implicit notion of resource \emph{ownership} (tracked via numeric values in a map, here) is a native notion in our specification methodology, explained in \secref{resources}. This contract's implementation enforces this intention by reverting if it is asked to transfer tokens away from anyone except the caller. Similarly, the right to mint \emph{new} tokens is restricted to a special privileged contract (represented by its \emph{address}), $\self.\code{minter}$.

The token contract can also be used to illustrate the infamous concept of \emph{re-entrancy vulnerabilities}: the subtle potential for a called contract to perform malicious callbacks and achieve undesirable outcomes. Say, for example, that lines 8 and 9 in the token contract were swapped, \ie{} the contract first called the receiver contract to notify it that it has received tokens, \emph{before} reducing their balance. If the notified contract called the sender of the transaction, it could in turn call back into the token contract and transfer the tokens it just transferred away \emph{a second time}; in particular, the \code{assert} on line~5 would not prevent the transfer because the balance was not yet updated at the time the callback happens. This would allow clients of the token contract to create tokens out of thin air. Variations of this pattern are behind most re-entrancy vulnerabilities, \eg{} the infamous DAO exploit~\cite{dao}; as we will show in \secref{resources}, our explicit resource reasoning will uncover such coding errors by default.

\section{Verification in the Presence of Unverified Code and Re-entrancy} \label{sec:unverified}

\begin{figure}
\includegraphics[width=0.8\textwidth]{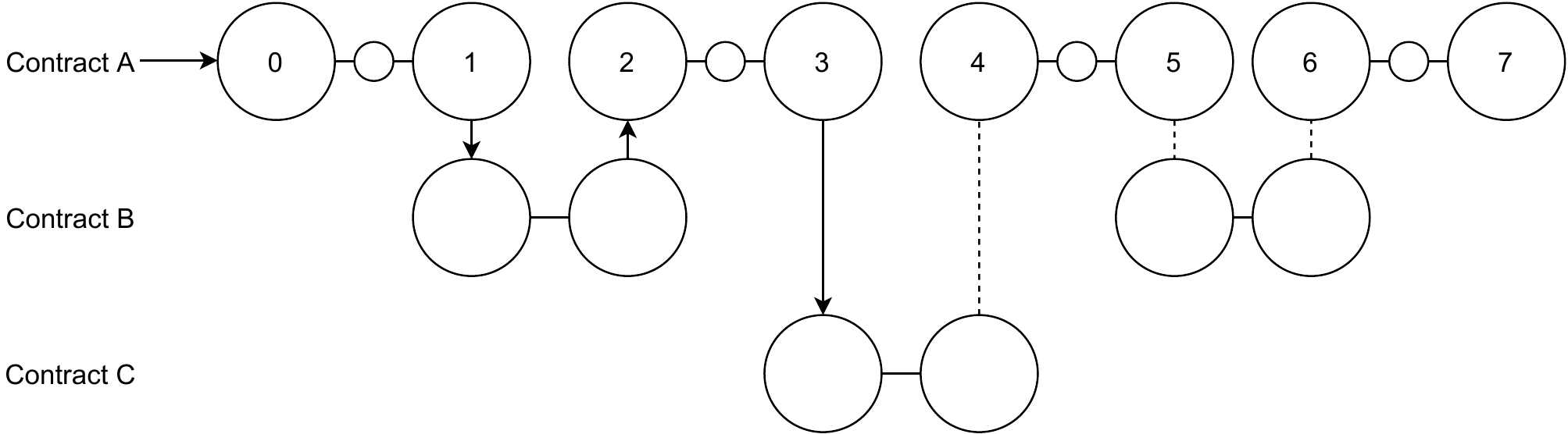}
\caption{Example of a smart contract transaction involving re-entrancy. Calls and returns are marked by arrows and dashed lines, respectively.}
\label{fig:public}
\end{figure}

Smart contracts frequently interact with other contracts; in particular, contracts that offer services to arbitrary clients often call functions on arbitrary other contracts. For example, the auction contract above sends Ether to (\ie{} calls) an arbitrary \code{msg.sender} in its \code{withdraw} function, which is necessary to ensure that any contract that has previously placed a bid and has been outbid can get back the Ether they sent.
While calls to functions of the same contract (\emph{internal} calls) can simply be verified by inlining the callee function, calls to other contracts (\emph{external} calls) are challenging for two reasons. First, as we explained earlier, the implementations of other contracts are in general not verified, and so we cannot reason about external calls using \eg{} standard pre- and postconditions.
Second, the implementations of other contracts are in general not known: we cannot make any assumptions about the callbacks they perform directly or via other contracts\footnote{It is common to limit the amount of gas sent with a call so that it is not sufficient to perform callbacks, but relying on this is considered bad practice, since the gas cost of Ethereum Virtual Machine instructions can change.}.
That is, we do not know if an external call simply modifies the state of the called contract and then returns, or if it triggers more complicated interactions such as those shown in Fig.~\ref{fig:public}: Contract $A$ calls contract $B$, which performs a re-entrant call to $A$; subsequently, $A$ performs an external call to a third contract $C$ before all calls return. In this scenario, when $A$'s call to $B$ returns, its own state may have changed as a result of the re-entrant call from $B$. Consequently, if the implementation of $B$ is unknown, one does not know which functions of $A$ have been re-entrantly called (if any), and how $A$'s state has changed as a result.
In this section, we present a specification and verification technique that is sound in the presence of unverified code and arbitrary re-entrancy. In the following, we assume that all calls are external (internal calls are inlined) and that the implementation of the callee is neither known nor verified (known or verified callees could provide stronger assumptions about the effects of the calls, but we focus on the common, most difficult case here).

\subsection{The Challenge}

The problem of re-entrancy by itself is not specific to smart contracts; it can also occur, for example, between objects in  object-oriented programming languages.
When reasoning about programs in such languages, this problem is usually solved by constraining what a called function may assume about the state and which parts of the heap it may manipulate~\cite{DBLP:books/sp/Muller02,ownership,sl,kassiosDF}. However, these verification techniques require that \emph{all} executed code is verified (and therefore known to adhere to the rules of the verification technique). In particular, called functions are typically verified to \emph{only} cause side effects allowed by the verification technique. In a smart contract setting, however, external code is often unverified, potentially malicious, and cannot be soundly assumed to follow \emph{any} particular rules beyond those of the execution environment. For the same reason, classical \emph{preconditions} on public functions are of limited use in this setting, since one cannot rely on external callers actually respecting them. In order to reason soundly, we have to conservatively assume that any external call  may lead to arbitrary callbacks into the original contract and, in particular, mutate the original contract's state in any possible way.

Some existing reasoning techniques for smart contracts either assume~\cite{verx} or aim to prove~\cite{DBLP:journals/pacmpl/GrossmanAGMRSZ18,DBLP:journals/pacmpl/AlbertGRRRS20} that re-entrancy cannot lead to behaviours that cannot also occur without re-entrancy (\ie{} that contracts are \emph{effectively callback-free} (ECF)~\cite{DBLP:journals/pacmpl/GrossmanAGMRSZ18}).
However, these techniques do not apply to the increasing number of contracts that use re-entrant calls as an essential part of their intended workflow (\eg{}~\cite{ERC1363}, which we explain \secref{evaluation}). In contrast, our methodology applies to \emph{all} contracts, even if they are not ECF\@:
Its central feature is that it allows users to express and prove critical properties of a contract despite the potential for arbitrary re-entrancy, as we explain next.

\subsection{Specification and Verification Technique}

We propose to use a two-pronged approach to smart contract specification and verification: (1)~We introduce a novel specification construct that lets us specify constraints on how a contract directly manipulates its own state. These constraints can be verified without considering external calls. (2)~We introduce two additional specification constructs that allow us to reason about external calls, even when the callee functions are unverified and potentially trigger re-entrant calls.
Both ingredients exploit a key feature of smart contracts: \emph{All} contract state is private, \ie{} cannot be directly modified by functions in other smart contracts. In particular, all updates to a contract's local state are performed by \emph{some} function from the contract's own code.

This feature provides a key distinction between smart contracts and most standard object-oriented languages; however, there are some non-smart contract languages (e.g. actors in Swift~\cite{swift}) that do offer this feature and can therefore also be verified using our approach. Generally, our technique enables code verification in the presence of re-entrancy and unverified code in any language that offers
(1)~objects (or contracts) with only object-private state, that is, the state of an object can be modified only by methods of that object, and
(2)~non-aliasing guarantees for object-private state, that is, there are never any mutable outside references to private object state.

\subsubsection*{Reasoning about call-free code.}

Since the state of a contract can be modified only by its own functions, one can express many important properties as constraints on local code, that is, the call-free code segments between (external) calls. For instance, if \emph{all} such code segments of a contract only ever increase the value of a counter, its value will never get smaller, even when external, potentially re-entrant functions are called. We refer to call-free sequences of statements as \emph{local segments}. In a sense, they represent the atomic operations a contract can perform: Outsiders can observe a contract's state \emph{between} local segments, but never in the middle of a segment. In Fig.~\ref{fig:public}, the local segments of contract A are the ones between the state pairs (0, 1), (2, 3), (4, 5) and (6, 7).

One class of properties that can be enforced by imposing constraints only on local segments is access control, \ie{} restricting the right to perform certain operations or modify certain data (indirectly, by calling a function on the contract) to specific callers. Access control is particularly important for smart contracts, since, unlike in standard object-oriented programs, they have to store \emph{all} the data of their clients in their own storage (\eg{} the balances in the token contract), making it vital to enforce that clients can only modify parts of that storage that conceptually belong to them. Access control restrictions are therefore a necessary part of the public specification of a contract. For example, for the auction contract from Fig.~\ref{fig:exampleauction} we may want to prove  that only the beneficiary can end the auction (by setting the \code{ended} flag).

To express constraints on local segments,
we introduce \emph{segment constraints}---two-state assertions (\ie{} assertions that can refer both to the current state and an \code{old} state) on the local state of a contract that must hold between the start and end states of \emph{each} local segment in the contract. Segment constraints are specified per contract, not per individual segment. In our auction example, we can express the access restriction for the \code{ended} field using the segment constraint $\code{msg.sender} \neq \old{\code{self.beneficiary}} \Rightarrow \code{self.ended} = \old{\code{self.ended}}$. Since, by definition, there are no external calls between the start and end states of a local segment, segment constraints are verified without considering external, unverified code. \figref{segconstrexample} illustrates the proof obligations generated by this example constraint for the \code{end} function from our auction contract.

\begin{figure}[t]
\begin{center}
\begin{minipage}[t]{0.8\textwidth}
\begin{lstlisting}
def end():
  @[\textcolor{blue}{$\{$ msg.sender $\neq$ self.beneficiary $\wedge$ self.ended = $x_1$ $\}$}@]
  assert not self.lock and not self.ended and msg.sender == self.beneficiary
  self.ended = True
  self.lock = True
  @[\textcolor{blue}{$\{$ self.ended = $x_1$ $\}$}@]
  @[\textcolor{gray}{ send(self.beneficiary, value=self.highestBid) }@]
  @[\textcolor{blue}{$\{$ msg.sender $\neq$ self.beneficiary $\wedge$ self.ended = $x_2$ $\}$}@]
  self.lock = False
  self.highestBid = 0
  @[\textcolor{blue}{$\{$ self.ended = $x_2$ $\}$}@]
\end{lstlisting}
\end{minipage}
\end{center}
\caption{Proof obligations generated by segment constraint $\code{msg.sender} \neq \mathit{old}(\code{self.beneficiary}) \Rightarrow \code{self.ended} = \mathit{old}(\code{self.ended})$ for function \code{end}. The function is divided into two local segments by the call in the middle; each segment is call-free and can therefore be verified using standard techniques. We use the logical variables $x_1$ and $x_2$ to represent the old values of the \code{ended} field. The same property has to be proved for every local segment in all other functions in the contract.} \label{fig:segconstrexample}
\end{figure}

\subsubsection*{Reasoning about (external) calls}

An external call may modify the state of the calling contract $A$ only via one or several re-entrant calls.
These re-entrant calls perform the modifications of $A$'s state by executing an arbitrary number of $A$'s functions, which in turn will execute some number of $A$'s local segments (in Fig.~\ref{fig:public}, the segments (2, 3) and (4, 5)).
Consequently, the reflexive and transitive closure of constraints describing the effects of $A$'s functions and segments can be used to soundly approximate the effects of an external call.
In the following, we introduce two complementary forms of such transitive constraints, which are useful for expressing different kinds of common properties. Both are \emph{auxiliary} specifications in the sense that they do not directly express vital correctness properties (unlike \eg{} segment constraints prescribing access control properties), but instead allow us to preserve properties across external calls.

\myparagraph{Transitive segment constraints.}
A vital property the \code{end} function of the auction contract must fulfill is that, when it returns, the contract's \code{ended} flag is set. Proving this requires showing that any re-entrant calls resulting from the \code{send}-command do not set the flag to \code{false}. We can show that this is the case because no local segment of the contract ever unsets the flag once it has been set, a property which can be expressed as a segment constraint.

The reflexive and transitive closure of all segment constraints of a contract describes the effect of an arbitrary number of local segments. Thus, it is known to hold between the pre-state and the post-state of any external call (\ie{} between states 1 and 6 in Fig.~\ref{fig:public}) and can be used to reason about such calls. Since it is generally not possible to compute the reflexive, transitive closure of our segment constraints automatically, we allow programmers to specify the \emph{transitive segment constraints} of a contract. These are checked to be reflexive and transitive, and are verified to hold across each local segment of the contract. Like segment constraints, transitive segment constraints are two-state assertions on the local state of a contract, similar to history constraints~\cite{liskov}. Since \emph{any} sequence of local segments is guaranteed to satisfy the transitive segment constraints of a contract, they may soundly be assumed to hold between the pre- and poststate of each external call.
Note that transitive segment constraints do \emph{not} subsume ordinary segment constraints, since  typical segment constraints used for access control (\eg{} the restriction on who can end an auction, shown above) are not transitive.

Transitive segment constraints are useful to express constancy properties, such as the fact that the auction's beneficiary never changes, or monotonicity properties like that of the \code{ended} field discussed above. The latter can be expressed using the transitive segment constraint $\old{\code{self.ended}} \Rightarrow \code{self.ended}$;  \figref{trsegconstrexample} illustrates how this constraint can be used to prove the desired postcondition for function \code{end} (even without the lock, which we will discuss below).

\begin{figure}[t]
\begin{center}
\begin{minipage}[t]{0.8\textwidth}
\begin{lstlisting}
def end():
  @[\textcolor{blue}{$\{$ self.ended = $x_1$ $\}$}@]
  assert not self.lock and not self.ended and msg.sender == self.beneficiary
  self.ended = True
  self.lock = True
  @[\textcolor{blue}{$\{$ ($x_1$ $\Rightarrow$ self.ended) }@]@[\textcolor{darkgreen}{$\wedge$ $x_3$ = self.ended}@] @[\textcolor{blue}{$\}$}@]
  @[\textcolor{gray}{ send(self.beneficiary, value=self.highestBid) }@]
  @[\textcolor{blue}{$\{$ self.ended = $x_2$}@] @[\textcolor{darkgreen}{$\wedge$ ($x_3$ $\Rightarrow$ self.ended)}@]@[\textcolor{blue}{$\}$}@]
  self.lock = False
  self.highestBid = 0
  @[\textcolor{blue}{$\{$ $x_2$ $\Rightarrow$ self.ended}@] @[\textcolor{red}{$\wedge$ self.ended $\}$}@]@[\textcolor{blue}{$\}$}@]
\end{lstlisting}
\end{minipage}
\end{center}
\caption{Proof outline showing the use of the transitive segment constraint $\mathit{old}(\code{self.ended}) \Rightarrow \code{self.ended}$ to prove the postcondition $\code{self.ended}$. We use logical variables $x_1$ and $x_2$ to represent the values of the \code{ended} field at the beginning of each local segment, and $x_3$ to represent its value before the \code{send} command.
The transitive segment constraint must be proved for every local segment of all functions (\textcolor{blue}{blue}). It can then be assumed to hold between the pre- and post-states of the external call (\textcolor{darkgreen}{green}), i.e., if \code{self.ended} is true before the call, we may assume it is true after the call. This is sufficient to prove the desired postcondition (\textcolor{red}{red}).} \label{fig:trsegconstrexample}
\end{figure}

Transitive segment constraints subsume single-state \emph{contract invariants}, which are often useful to specify consistency conditions on contract states, which must hold whenever the contract relinquishes  control to other contracts (and, thus, its state becomes observable to the environment). The verification of transitive segment constraints implies that single-state contract invariants hold at the end of each local segment, which includes the state before any call as well as the post-state of each function. Consequently, each function may soundly assume such contract invariants to hold in its pre-state, as well as after the return of each external call, analogously to class or object invariants in object-oriented programs~\cite{Leavens07,DrossopoulouFrancalanzaMuellerSummers08}.
For the auction, an important invariant is that its funds suffice to pay all its obligations, which can be written as the transitive segment constraint $\code{self.balance} \geq \mathit{sum}(\code{self.pendingReturns}) + \code{self.highestBid}$.

\myparagraph{Function constraints.}
It is common that each individual function of a contract \emph{as a whole} satisfies a two-state property, even if some of its local segments do not. Such situations occur for instance if \emph{some} sequences of local segments violate the property, but no function in the contract ever executes such a sequence. The re-entrancy lock in the auction contract is an example: The field \code{self.lock} is set to true by \code{withdraw} and \code{end} before their calls to \code{send}, and reset to false afterwards. Since each function of the contract reverts if the locks is set, this pattern ensures that each function of the auction contract leaves the contract state completely unchanged if the lock is set in its pre-state.

However, this property cannot be verified as a transitive segment constraint: Some local segments reset the lock, such that any subsequent state change violates the property. That is, the property does not hold for arbitrary sequences of local segments, but it does hold between the pre-state and the post-state of each contract function. Note that any external call can modify the contract state only by executing these contract functions (via re-entrant calls) from start to finish.

To exploit this fact, we introduce \emph{function constraints}: two-state assertions on the local state of a contract that must hold between the pre- and post-state of every function in the contract. In Fig.~\ref{fig:public}, this means they have to hold between states~2 and~5 as well as states~0 and~7.
Like transitive segment constraints, function constraints are specified per contract; they must be satisfied by \emph{all} of its functions (reflecting that we do not know statically which re-entrant calls are triggered by an external call). Since external calls may trigger the execution of an arbitrary number of contract functions, we require function constraints to be reflexive and transitive.
For the lock example, we can express the desired property as the function constraint $\old{\code{self.lock}} \Rightarrow \code{self} = \old{\code{self}}$.

Note that function constraints do \emph{not} subsume transitive segment constraints. For instance, in the special case of single-state assertions, transitive segment constraints (that is, the contract invariants discussed above) are known to hold before each call and may, thus, be assumed in the pre-state of each function, whereas function constraints may not. Neither do transitive segment constraints subsume function constraints, as we illustrated with the lock example above.
With these two specification constructs, we can modularly verify properties in the presence of calls to unverified contracts with arbitrary re-entrancy.
In \secref{resources}, we will complement these constraints with effect specifications on a contract's resources to obtain even stronger guarantees.

\section{Inter-Contract Invariants}\label{sec:ici}

\begin{figure}
\begin{center}
\begin{minipage}[t]{0.75\textwidth}
\begin{lstlisting}
interface Token:
  balances: map(address, uint256)

  def transfer(from: address, to: address, amount: uint256):
    pass

contract Auction:
  token: Token

  def withdraw():
    assert not self.lock
    toSend = self.pendingReturns[msg.sender]
    self.pendingReturns[msg.sender] = 0
    self.lock = True
    self.token.transfer(self, msg.sender, toSend)
    self.lock = False

  def distributeExcess():
    excess: int128 = self.token.balances[self] - sum(self.pendingReturns)
    excess -= self.highestBid
    assert excess != 0
    perBidder: int128 = excess / size(self.pendingReturns)
    for bidder in keys(self.pendingReturns):
      self.pendingReturns[bidder] += perBidder
\end{lstlisting}
\end{minipage}
\end{center}
\caption{Minimal interface of the token contract in Fig.~\ref{fig:exampletoken}, and part of an adapted auction contract that deals in tokens and additionally has a function that distributes excess tokens among previous bidders.} \label{fig:interface}
\end{figure}

Smart contract applications are frequently implemented via multiple contracts which call one another. As in many programming languages, the \emph{interfaces} of the Vyper and Solidity languages are designed to facilitate such collaborations. Interfaces declare that a contract offers \emph{at least} some set of functions and fields\footnote{Interfaces actually contain \code{constant} functions that guarantee not to modify any state instead of fields, but we model them as fields here to simplify the presentation.}, but do not give any information about their implementation, or preclude the existence of additional functions in the contract. Therefore, they decouple client contracts (in the software engineering sense) from the concrete implementations of the contracts they build on.

For example, an auction contract similar to our example from \figref{exampleauction} could instead deal in tokens conforming to \eg{} the ERC20 standard interface~\cite{ERC20}. Fig.~\ref{fig:interface} shows a minimal interface of the token contract from Fig.~\ref{fig:exampletoken} as well as (part of) a modified version of the auction contract, where calls to \code{send} are replaced by calls to the token contract's \code{transfer} function (we will discuss the added function \code{distributeExcess} later).

However, our techniques for equipping contracts with invariants and proof obligations to maintain them no-longer suffice for collaborating contracts, since such collaborations naturally give rise to invariants that depend on the state of \emph{other} contracts. For example, the modified auction still needs an invariant that it has sufficient funds (now tokens) to pay its obligations to all participants, which can be expressed in terms of the states of \emph{both} the auction and token contract by: $\code{self.highestBid} + \mathit{sum}(\code{self.pendingReturns}) \geq \code{self.token.balances[self]}$.
In this section, we extend the technique presented in \secref{unverified} to such \emph{inter-contract invariants}\footnote{We focus on single-state invariants here for simplicity only: our technical solution also supports two-state assertions.}.

An inter-contract invariant has a single \emph{primary} contract (the contract depending directly on the property); any other contracts whose state is mentioned are its \emph{secondary} contracts.
In the example, the modified auction contract is the primary contract, since the auction's funds must be sufficient for the auction to function correctly, whereas the token contract can have many (non-auction) clients with different functionality and is not responsible for their correctness. This asymmetry is reflected in the above invariant, where \code{self} is the auction contract.

\begin{figure}
\begin{center}
\begin{minipage}[t]{0.65\textwidth}
\begin{lstlisting}
contract BadToken implements Token:

  def steal(from: address, amount: uint256):
    assert self.balances[from] >= amount
    self.balances[msg.sender] += amount
    self.balances[from] -= amount

  def transfer(from: address, to: address, amount: uint256):
    assert self.balances[from] >= amount and msg.sender == from
    newAmount: uint256 = self.balances[from] - amount
    self.balances[to] += amount
    self.balances[from] = 0
    thirdparty.notify()
    assert self.balances[from] == 0
    self.balances[from] = newAmount

contract ThirdParty:
  auction: Auction

  def notify():
    auction.distributeExcess()
\end{lstlisting}
\end{minipage}
\end{center}
\caption{Possible implementation of the \code{Token} interface. The implementing contract may offer additional functions, \eg{} in this case, one that allows anyone to steal tokens from any existing account.} \label{fig:badinterfaceimpl}
\end{figure}

Ensuring non-trivial inter-contract invariants requires that
both the primary and all secondary contracts are verified; the state of unverified contracts may change arbitrarily, which precludes the verification of invariants that depend on it. However, all contracts other than the primary and secondary ones may still be unverified, and as before, verified contracts may still call functions of unverified ones.
Additionally, we do \emph{not} depend on having access to the implementations of the secondary contracts. These may in particular be hidden behind interfaces.

\subsection{Challenges}

Modular verification of inter-contract invariants poses two main challenges:

\myparagraph{Challenge 1: Missing encapsulation.}
The first challenge is that the state that an inter-contract invariant depends on is not fully-encapsulated in the way we have exploited so far: It is now no longer the case that the invariant can only be broken by code of the primary contract; instead, it can be also be broken by the code of a secondary contract, which may not be known.

To illustrate this challenge, consider a scenario in which the token contract has a function \code{steal} that lets an arbitrary contract steal another contract's funds, as shown in \figref{badinterfaceimpl}: If this function existed, a third party could call it to steal the tokens of the auction contract; if the auction contract also has any pending returns, this would break our inter-contract invariant. Note that there is nothing the primary contract can do to prevent this; in fact, this can even happen in a transaction that does not involve the primary contract at all. Additionally, one cannot conclude from the token contract's interface alone whether or not it has such a function (or any other function that allows one contract to decrease another contract's funds).

\myparagraph{Challenge 2: Temporarily-broken invariants.}
Second, secondary contracts may \emph{temporarily} break inter-contract invariants when called by the primary contract in a way that makes the inconsistent state visible to other contracts. Note that it is normal and unavoidable that invariants are temporarily broken; however, states in which this is the case must never be visible to outside contracts, which can be the case here.
This challenge is illustrated in function \code{transfer} in \figref{badinterfaceimpl}. This function performs a token transfer from one contract to another, as it should, but (perhaps as a clumsy attempt to avoid a DAO-like re-entrancy vulnerability) it temporarily sets the balance of the token sender to zero and performs a call to the outside, before restoring the balances to the desired end state.
This can lead to problematic behaviour: Assume that the auction contract has non-zero pending returns for two contracts, A and B, and that contract B is the \code{ThirdParty} contract shown in \figref{badinterfaceimpl}.
If contract A calls \code{withdraw}, the auction contract will call \code{transfer}, which will set its token balance to zero. Now the token contract calls function \code{notify} of contract B. Note that, in this state, the inter-contract invariant is broken: The auction contract still has pending returns for B, but its current balance is zero.
When B in turn calls the function \code{distributeExcess} in the auction contract this state, this function does not work as designed: The purpose of the function is to distribute any excess tokens the auction contract may own among previous bidders (which may be part of some intended functionality where third parties are rewarded for taking part in the auction, or simply a failsafe in case someone accidentally transfers tokens to the auction contract). It calculates the excess by subtracting the sum of the pending returns and the current highest bid from the auction contract's token balance, assuming that the result will be non-negative, which \emph{should} be guaranteed by the invariant. As a result, since we assume the invariants (transitive segment constraints) at the beginning of each function, we could prove a postcondition here that states that pending returns can only be increased by this function.
Now, however, the result can actually be negative, and as a result, the pending returns of all previous bidders will be decreased, breaking the postcondition.
We must therefore adjust our verification technique to ensure that this invariant cannot be proved for this contract, since it does not hold in practice.

Note that, again, it is not possible to see from the interface that this problem exists: If function \code{transfer} has a postcondition that describes its behaviour, it will state that (by the time the function returns) the transfer has been executed as desired; nor is it possible to see from an interface whether the function performs any calls to the outside. Also note that, unlike the previous problem, this problem is unrelated to encapsulation: Now, it is not third parties that can modify state in unintended ways, but it is the primary contract itself (which \emph{must} be able to modify the state in the token contract that conceptually belongs to it) whose call has unintended consequences.%

\subsection{Solution}

In order to enable modular verification of inter-contract invariants, we build on our existing approach for proving transitive segment constraints, introduce one additional specification construct, and add proof obligations that prevent both of the potential problems mentioned above.
More concretely, transitive segment constraints are now (unlike all other specification constructs we introduced) allowed to express inter-contract invariants, \ie{} they may now refer to the state of other contracts that are reachable from the primary contract.
All existing proof obligations for transitive segment constraints remain, that is, we check that they are reflexive and transitive, prove that they are established by the primary contract's constructor, and verify that each local segment of the primary contract satisfies them.
Additionally, interfaces may declare both function postconditions and transitive segment constraints, which all contracts implementing the interface must adhere to.

We now address Challenge 1 by extending interfaces with specifications that provide the missing guarantees: we allow annotating interfaces with novel
\emph{privacy constraints}, which express which part of the contract state conceptually belongs to the caller of a function and, thus, cannot be freely manipulated by other callers. Essentially, a privacy constraint extends the encapsulation guarantees that already exist for the state of the primary contract to (parts of) the state of a secondary contract.
Privacy constraints are segment constraints of the form $\forall a . \msg.\code{sender} \neq a \Rightarrow P$, where $P$ is reflexive and transitive. The privacy constraints specified in an interface must be satisfied by \emph{all} functions of a contract implementing the interface, even those not mentioned in the interface.

The privacy constraint $\forall a . \msg.\code{sender} \neq a \Rightarrow \code{self.balances[a]} \geq \old{\code{self.balances[a]}}$ on the token interface from Fig.~\ref{fig:interface} expresses that a caller may increase the balance of any contract, but decrease only its own. Since the \code{steal} function from \figref{badinterfaceimpl} violates this property, \code{BadToken} is now no longer a valid implementation of the \code{Token} interface. In other words, the privacy constraint constitutes a promise that the secondary contract does not contain \emph{any} function that will allow third parties to decrease the auction contract's balance.
This is exactly the information needed to prove that calls on the token contract by third parties cannot violate the inter-contract invariant stated at the beginning of this section.

In general, assuming that all secondary contracts are annotated with privacy constraints, we prove that those privacy constraints re-encapsulate the state our invariant depends on as follows: 
We require that each inter-contract invariant
is \emph{stable} under state changes allowed by the privacy constraints of all secondary contracts, 
\ie{} that the privacy constraints forbid all changes that could break the invariant. %
We formally define the notion of assertion stability in Sec.~\ref{sec:logic} and illustrate it here with our example:
Assume that a function of the secondary token contract is called by a party other than the primary contract. For any local segment of the token contract,
the token's privacy constraint shown above guarantees that $\code{self.token.balances[self]} \geq \old{\code{self.token.balances[self]}}$. 
If, in the old state, the inter-contract invariant $\code{self.highestBid} + \mathit{sum}(\code{self.pendingReturns}) \geq \code{self.token.balances[self]}$ held, then the privacy constraint 
(along with the knowledge that the local segment of the secondary contract cannot directly change the state of any other contracts) implies that the inter-contract invariant also holds in the new state.

Challenge 2 is not addressed by the introduction of privacy constraints; since the caller in this scenario is the primary contract itself, privacy constraints offer no guarantees about the potential behaviour of the secondary contract.
To address the second challenge, we require that, in every state where the primary contract calls a secondary contract, \emph{any} changes a local segment of the secondary contract can make cannot break the invariant. That is, the conjunction of the privacy constraints and transitive segment constraints of all \emph{other} secondary contracts, \emph{excluding} the called one, conjoined with information we have about the primary contract, must suffice to show that the invariant cannot be broken by the called secondary contract in the state where the call is made.
This criterion can again be expressed as a stability constraint (as shown in \fullversionref{formalisation}).

For the invariant we attempted to prove, this criterion is not fulfilled, since, as we showed, the secondary contract can break the invariant.
We can, however, prove a weaker invariant that still serves our purpose: 
we exploit that whenever both contracts' state may be out-of-sync, the \code{lock} is set and the auction contract cannot be called. 
So, we require that the desired consistency criterion (that the auction has sufficient funds to pay its obligations) is true whenever the \code{lock} field is not set\footnote{This pattern is often used in OO-verification when proving invariants between multiple objects~\cite{LeinoMueller04}.}, resulting in the invariant
$\neg\code{self.lock} \Rightarrow \code{self.highestBid} + \mathit{sum}(\code{self.pendingReturns}) \geq \code{self.token.balances[self]}$.

This invariant actually holds, and  makes explicit that \code{distributeExcess} can rely on its funds being greater or equal the contract's obligations \emph{only} when the lock is not set. As a result, it will now be impossible to prove a postcondition for this function stating that it only increases pending returns, unless the function is fixed by adding \code{assert not self.lock} at the beginning. 

We can prove our adapted invariant as follows:
We show that whenever the primary contract calls the secondary contract, the assertion \code{self.lock} holds, which implies that our adapted invariant holds as well. This assertion is independent of the token contract and so cannot be broken by changes to its state, and it is preserved by calls to the primary contract (which we can prove using a suitable function constraint $\old{\code{self.lock}} \Rightarrow \code{self.lock}$); that is, it fulfills our stability criterion.

The proof obligations we have outlined ensure that secondary contracts cannot break inter-contract invariants when called by the primary contract (Challenge 2) or anyone else (Challenge 1). Along with the proof obligations that ensure that the primary contract establishes and maintains the invariant, which we described in the previous section, this is sufficient to guarantee that the inter-contract invariant will hold at the end of every local segment of any contract.
In summary, the added proof obligations generalise our previously-introduced specification constructs to verify invariants of collaborating contracts. This verification is fully modular, based on the implementation of the primary contract and specified interfaces for all secondary contracts. It is sound even when these contracts interact with unverified code, and in the presence of arbitrary re-entrancy.

\section{Resource-Based Specifications} \label{sec:resources}

The vast majority of smart contracts in some way model resources and resource transfers, such as the token and auction contracts we have seen before.
Resources have a number of basic properties that are important for the correctness of every contract that works with them: they \emph{cannot be duplicated}, they \emph{have an owner}, and they \emph{cannot be taken away from that owner without their consent}. Explicitly specifying these properties for every smart contract that uses some sort of resource is possible, but laborious and error-prone.
Instead, we propose a dedicated specification and verification technique that has basic resource properties built-in and that offers high-level specification constructs to declare resources and to describe resource transactions.
The potential of resource-based reasoning for smart contracts has been recognized before; for instance, Move~\cite{move} has native support for resources in the blockchain and language (but does not have built-in guarantees of all the basic properties mentioned above).
Compared to specifications that express resource properties via changes of the contract state,
our resource specification system has three main advantages
(note that Move, due to it having a different resource model, does not have these three advantages built-in, see Sec.~\ref{sec:related}):

\begin{enumerate}
  \item Safety:
  Basic properties of resources, such as the fact that they cannot be duplicated and cannot be taken away from their current owner without their consent, are baked into the system.
  Our verification approach ensures that these properties hold by default, without developers having to specify them, so that there is no danger that important properties are missing in the specifications, and there is no need to write them down for every contract.

  \item Higher-level reasoning:
Developers think about resources as an abstract concept; for instance, they think of a token as a kind of currency, not some contract whose state contains a map. Resource-based specifications let developers describe their contracts' states and interactions on this abstraction level, leading to simpler and more intuitive specifications.

  \item Client documentation:
Writing postcondition-based specifications for smart contract functions is often difficult because of potentially re-entrant calls with unbounded effects. Our resource system enables users to prove novel effect-based function specifications that give a caller an upper bound on the negative consequences it may suffer from calling a function (\eg{} losing some Ether) and a lower bound on the positive consequences (\eg{} receiving some tokens).

\end{enumerate}

In this section, we describe the basic attributes of our resources, the operations that can be performed on them, and how we connect the contract's actual state to the resource state. We show how effect-based function specifications based on resources give callers extra information. Finally, we describe advanced concepts such as \emph{derived} resources, representing titles to other resources.

\subsection{Resource Model}

\begin{figure}
\includegraphics[width=0.8\textwidth]{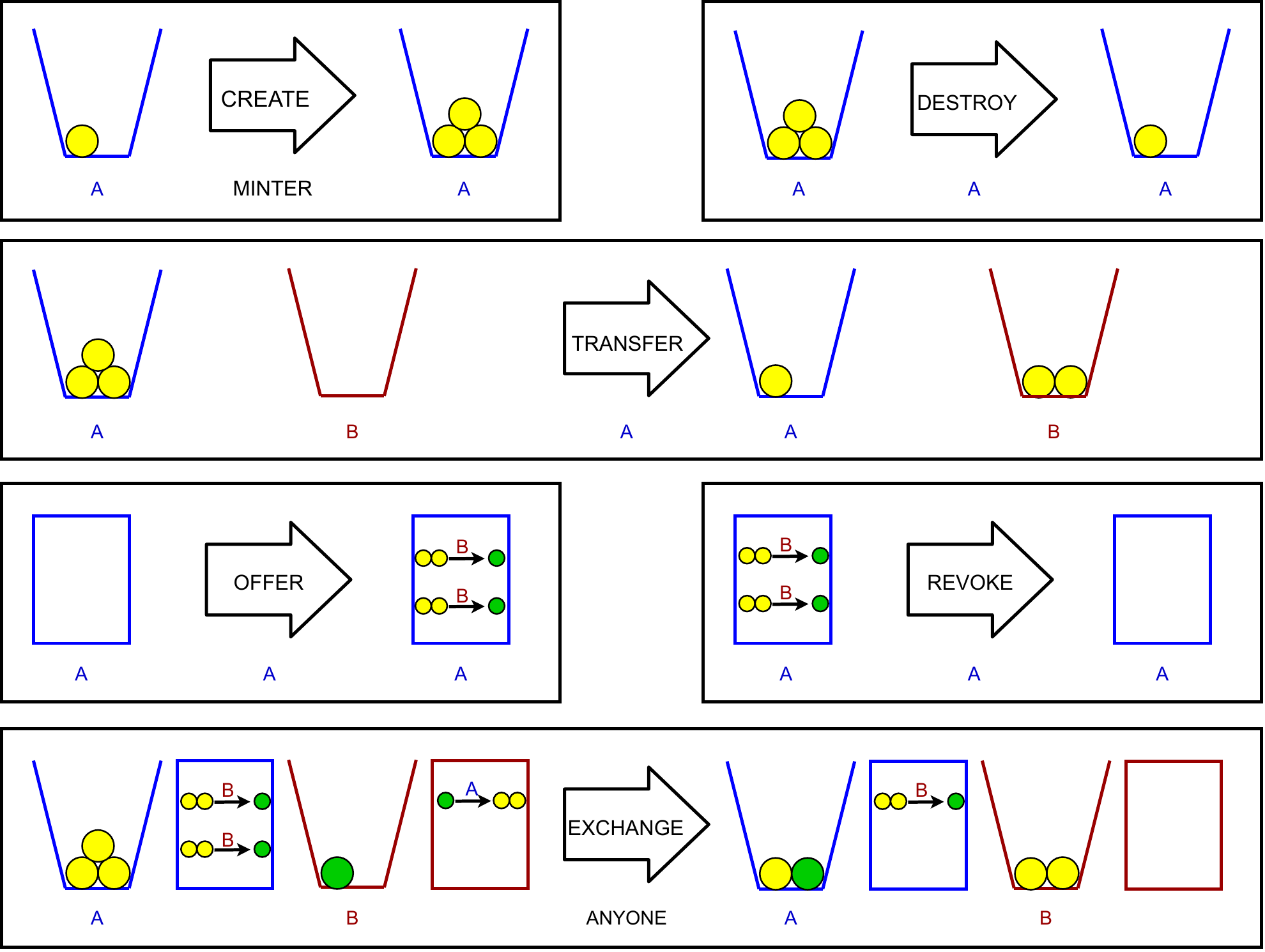}
\caption{Operations on resources and offers. Buckets represent resources owned by addresses $A$ and $B$; the rectangles contain the offers they have made. Big arrows show the name of the resource operation; the names under the arrows show who can perform the operation (create-operations may be performed by anyone who has a special resource representing the right to mint new resources). For example, only $A$ can \emph{transfer} two of its three yellow resources to $B$, or it can \emph{offer} $B$ to exchange two of its yellow resources against a green one.}\label{fig:resources}
\end{figure}

In our system, a resource can represent anything that cannot be duplicated, has an \emph{owner}, and has some non-negative value.
Resources are owned by addresses on the blockchain. Ownership implies control of the resource, \ie{} only the owner of a resource can transfer or destroy it.
Receiving additional resources does not require consent. In this regard, our resources are similar to Ethereum's built-in Ether resource (which is treated as a built-in resource in our system).

Fig.~\ref{fig:resources} shows the operations that can be performed on resources, and who is allowed to perform them. Resources can be \emph{created} by privileged parties who have the right to do so (usually called minters). They can be \emph{transferred} to other addresses or \emph{destroyed} by their owners (and by noone else).
In addition, addresses can make \emph{offers} to exchange some resources against others; when an offer from one address exists and a second party makes a matching counter-offer, the \emph{exchange} can be performed at an arbitrary point in time, without further agreement by the involved parties (consuming the offers). For maximum flexibility, an address need not own the resources it offers at the point when it makes the offer, but an exchange requires that both addresses actually hold the offered resources.
Addresses can \emph{revoke} previous offers they have made. The resulting set of operations is simple but sufficient to model the behaviour of a wide range of real smart contracts.

\subsection{Resource State}
Every smart contract may declare one or more resources that they implement (\eg{} token contracts would declare a token resource). For each resource, all addresses implicitly have a balance (as for the built-in Ether). Similarly, each address has a set of existing offers on the resources it declares. These balances and offer sets are \emph{ghost state}: state that exists only for verification purposes, but is not present at execution time. Our specifications can refer to this state, \eg{} a postcondition can refer to the caller's balance for resource $R$ as $\mallocated{R}[\code{msg.sender}]$. Note that specifications about resource state can be arbitrarily combined with the other specification constructs we have introduced; for example, one could write a segment constraint stating that a contract may perform some operation only if it owns some minimum amount of a resource.

The resource ghost state can be changed only by executing \emph{ghost commands}, written in the verified contract, that each perform one of the resource operations mentioned above. As an example, the ghost command $\creallocate{R}{f}{t}{a}$ transfers $a$ amount of resource $R$ from $f$ to $t$. This ghost command requires that $f$ has sufficient amounts of $R$, and that $f$ is the contract invoking the ghost command, \ie{} that $f$ has called the function that contains it (modulo delegation, which we discuss later). These conditions are checked by the verifier, and they enforce the basic properties of the resource system; the latter check in particular enforces ownership constraints.
The resource ghost state has the same encapsulation as ordinary contract state, that is, the ghost commands in a contract can modify only the state of the resources declared in that contract.
A more detailed description of the ghost commands from \figref{resources} is shown in \fullversionref{ghost-commands}.

\subsection{Connecting Resource State and Contract State}

In order to be useful for verification, the resource ghost state must be connected to the contract's actual state.
Our system achieves this by letting developers write invariants (\ie{} transitive segment constraints) that relate the resource ghost state (\ie{} balances and existing offers) to the contract state.
 When verifying a contract, we then enforce that these coupling invariants, like all transitive segment constraints, hold at the end of every local segment. For the token contract, the invariant would be $\mallocated{\mathit{token}} = \code{self.balances}$, meaning that the balances of the resource named ``token'' are recorded in the contract's \code{balances} field.

This check essentially forces changes on the resource state and on the contract state to happen in lockstep:
If a change happens on the resource state with no equivalent change on the contract state (or vice versa), the invariant cannot be verified.
As a result, the properties our system guarantees for resources (like ownership) carry over to the actual contract state. For instance, our system prevents the verification of a function \code{steal} that allows arbitrary callers to steal some client's funds: the modification to the contract state must be mirrored in a corresponding change of the resource state (by the coupling invariant). The ghost command that makes this change checks that the transferred funds belong to the function's caller, which would fail here.

\begin{figure}[t]
\begin{center}
\begin{minipage}[t]{0.7\textwidth}
\begin{lstlisting}
def transfer(from: address, to: address, amount: uint256):
  @[\textcolor{blue}{$\{$ balances$_{\mathit{token}}$ = self.balances  $\}$}@]
  assert self.balances[from] >= amount and msg.sender == from
  self.balances[to] += amount
  self.balances[from] -= amount
  @[\textcolor{red}{ transfer$_{\mathit{token}}$(from, to, amount) }@]
  @[\textcolor{blue}{$\{$ balances$_{\mathit{token}}$ = self.balances  $\}$}@]
  @[\textcolor{gray}{to.notify(from, self, amount)}@]
\end{lstlisting}
\end{minipage}
\end{center}
\caption{Example of contract state and resource state moving in lockstep. In order to re-establish the coupling invariant (\textcolor{blue}{blue}) at the end of the first local segment of the token contract's \code{transfer} function after modifying the contract state, one must execute a \code{transfer} command (\textcolor{red}{red}) that modifies the resource state accordingly.
Verification ensures that all conditions imposed by the resource model are fulfilled when \code{transfer} is executed.} \label{fig:ghostcommand}
\end{figure}

As an example, in the \code{transfer} function of the token contract, we would have to insert the ghost command $\creallocate{\mathit{token}}{\code{from}}{\code{to}}{\code{amount}}$ before the call to \code{notify} in order to re-establish the invariant, as shown in \figref{ghostcommand}.
Verifying the function requires proving that \code{from} is the  $\msg.\code{sender}$ and that \code{from} owns at least \code{amount} tokens, both of which we can prove because of the assertion at the beginning of the function.

\subsection{Client Specifications}

The system described so far guarantees that others cannot take away the resources owned by a contract.
However, the contract itself may perform operations that lead to a loss of resources, \eg{} by transferring or destroying them.
Our rules for resource commands ensure that any such operation is initiated by the owner calling a function on the contract that declares the resource (\eg{} a function on the token contract that contains a \code{transfer} ghost command). Therefore, it is vital that functions provide callers with specifications describing how they affect the caller's resources.

We address this problem by introducing \emph{effects-clauses} on contract functions, which specify ghost commands that will be executed when the function is called (assuming it does not revert).
Each function has a multiset of effects, and each effect corresponds directly to one of the ghost commands introduced above, meaning that there are effects for creating, transferring, and exchanging resources, etc. Effects-clauses are unordered and do not give any information regarding \emph{when} during the call the effects occur. As an example, if a function's effects-clause contains $\ereallocate{R}{\code{msg.sender}}{t}{4}$ and $\ereallocate{R}{\code{msg.sender}}{t}{6}$, this means that after successful execution, it will have transferred 10 $R$ (in two separate steps of 4 and 6) from the caller to address $t$ at some point during the call.

In contrast to traditional effects-systems, our effects-clauses are \emph{not} required to be transitive: the ghost operations performed directly by the called contract's function must be included, but those caused by further external calls made by this contract need not be tracked. This non-standard design is motivated by the presence of unverified code; ultimately, there will be cases in which we could not know the effects of arbitrary external code. The rules for checking effects clauses are simple: (a)~the effects-clause of a function \emph{must} contain the effects of all ghost commands directly in its body, and (b)~it \emph{may} declare any additional effects resulting from its own calls to other functions.

This might sound problematic: a caller of a function is only able to see the effects clauses (and postcondition) to judge whether they \emph{consent} to these effects happening, but if the effects do not track all transitive calls, one might expect that the caller could be tricked into allowing \eg{} more of their resource to be transferred than they realise. Perhaps surprisingly, due to the resource ownership principles built into our methodology, our non-transitive effects actually remain powerful and useful: they ultimately describe worst-case information about what could happen to a caller's resources by calling the function in question \emph{except possibly for additional calls that go via the same original caller.} In other words, the caller is explicitly consenting to at most these effects happening, unless they subsequently consent to additional effects by making a further call.

To see why this is the case, consider a function that may have a negative effect on a calling contract's resources. 
Since all ghost operations that have a negative effect impose a proof obligation that the resources are owned by \code{msg.sender}, the negative effect \emph{must} occur either in the initially-called function, or, after a sequence of additional calls, in some function that was again called \emph{by the original caller}. 
In the first case, according to check~(a), this effect will be included in the initially-called function's effect clause: the caller was aware of the effect and allowed it to happen by making the call.
In the second case, for the same reason, this effect must be declared on the subsequently-called function called by the same caller, who consents to the additional effects by making this subsequent call.
Note that it \emph{is} possible that a call will cause effects that are positive or neutral for the caller (\eg{} an unknown contract giving them tokens) which the called function did not declare; however, since those are not negative for the caller, not knowing about them does not impact the caller's ability to consent to the call.

As a result, our effects-clauses enable each contract to know which negative effects a call may have on its resources, such that it can refrain from making calls with undesired effects. This solution gives strong guarantees in the presence of arbitrary re-entrancy, when it is impossible to give the called function a precise postcondition.
\fullversionref{example_token} illustrates the use of effects-clauses %
on (an extended version of) the token contract from before.

\subsection{Derived Resources}

As a final ingredient, our system contains one additional concept to model the difference between physically having a resource and conceptually being its rightful owner.
As an example, consider the auction contract again: Whenever a bidder sends some wei to it, that wei now physically belongs to the auction contract, which could in principle do with it whatever it wants. However, conceptually, as long as the auction is running, the bidder is still the owner of the wei it sent, and it rightfully expects to either be able to get it back later (if someone else makes a higher bid) or to exchange it for the auctioned good when the auction ends. That is, after a bid and before the end of the auction, the physical owner of that wei (the auction contract) is different from the conceptual owner (the bidder). This is a relatively common notion that occurs whenever some contract (temporarily) manages another contract's resources, and it obviously comes with certain expectations (\eg{} the auction contract should not be able to give the wei it has to anyone but its rightful owners).

Our system has support for this scenario in the form of \emph{derived resources}, representing conceptual ownership of a resource physically owned by someone else; essentially, a kind of \emph{title}.
In our example, the auction contract could declare a resource \code{wei_in_auction} derived from the built-in wei resource, as shown in \figref{derivedexamplepart}. When a bidder sends wei to the auction contract by calling function \code{bid}, it transfers its wei to it, but gets the same amount of \code{wei_in_auction} in return, signifying that it is owed that amount of wei from the auction contract.
 If another higher bid comes in and the bidder gets its wei back by calling \code{withdraw}, its titles are destroyed again. In contrast, the winner of the auction exchanges its titles against the auctioned good, so that their bid is now owed to the beneficiary of the auction.
 At any given point, the amount of titles address $c$ has in the auction contract is $\code{self.pendingReturns[}c\code{]} + (\code{self.highestBidder} = c ? \code{self.highestBid} : 0)$, meaning that this is also the amount of \code{wei_in_auction} contract $c$ owns.

\myparagraph{Resource creation and destruction.} The existence of an instance of a derived resource is always bound to an instance of the resource it is derived from. That is, if a contract declares resource $D$ derived from another resource $R$, then an instance of $D$ comes into existence for every instance of $R$ it receives (via a transfer operation or an exchange), and is automatically allocated to the sender of the $R$; there is no way to create an instance of $D$ without receiving an instance of $R$. Similarly, whenever the contract sends some amount of $R$ to someone else, this destroys the same number of $D$ instances that other contract currently owns. This mechanism ensures that the contract always owns enough of the original resource to ``pay back'' its title loans. The reader may recall that this fact was an invariant of the auction contract that we explicitly mentioned in Sec.~\ref{sec:unverified}; now, with derived resources, this invariant is checked automatically and does not have to be specified explicitly.

\myparagraph{Resource transfers.} In order to ensure that contracts do not give away resources that (according to an existing title) belong to someone else, our system enforces that the contract may now transfer $R$ to another contract \emph{only} if that other contract already owns a sufficient amount of $D$, \ie{} the original contract already owes the second contract at least the amount to be transferred.
As an example, when the auction contract sends some amount of wei to a previous bidder of the auction in line 18, this is allowed only if the bidder currently owns an equal amount of $\code{wei_in_auction}$, and if the beneficiary has offered to exchange its $\code{wei_in_auction}$ back to ordinary wei. If this is the case, then, the moment the send executes, that amount of the beneficiary's $\code{wei_in_auction}$ is automatically destroyed, and the offer to exchange it is consumed.

Apart from the aforementioned restrictions, derived resources behave just like other resources. In particular, they can be traded like other resources (\eg{} someone could pay for some good in  \code{wei_in_auction}, meaning that they give the right to get wei back from the auction contract to someone else). This is relevant for some DeFi contracts that give out tokens that represent ownership of some other goods (\ie{} derived resources), but are traded as tokens on their own.

\myparagraph{Proof technique.} Our proof technique enforces the properties listed above by automatically creating and/or destroying instances of the derived resource whenever a contract calls an external function that declares (in its effects-clause) that it performs a transfer or exchange of the underlying resource to or from the calling contract. Sending or receiving wei is a special case but is treated analogously, \ie{} when sending wei, this is handled as if the called function declared that it transfers wei away from the calling contract.
To avoid that a contract loses resources that conceptually belong to others without its knowledge (which would mean that it cannot perform the aforementioned checks), our system enforces that the contract declaring $D$ cannot make offers to give away $R$, since such offers could result in the contract losing $R$-instances (when the exchange happens) at an arbitrary point in the future.

\fullversionref{example_auction} shows the entire auction contract with derived resource specifications.

\begin{figure}[t]
\begin{center}
\begin{minipage}[t]{0.75\textwidth}
\begin{lstlisting}
@[\textcolor{darkgreen}{resource: good() }@]
@[\textcolor{darkgreen}{resource: wei\_in\_auction() derived from wei }@]

@[\textcolor{darkgreen}{performs: create[wei\_in\_auction](msg.value) }@]
@[\textcolor{darkgreen}{performs: offer[wei\_in\_auction <-> good](msg.value, 1, to=self.beneficiary, times=1) }@]
def bid():
    assert block.timestamp < self.auctionEnd and not self.ended
    assert msg.value > self.highestBid and msg.sender != self.beneficiary
    @[\textcolor{red}{ offer[wei\_in\_auction <-> good](msg.value, 1, to=self.beneficiary, times=1) }@]
    self.pendingReturns[self.highestBidder] += self.highestBid
    self.highestBidder = msg.sender
    self.highestBid = msg.value

@[\textcolor{darkgreen}{performs: destroy[wei\_in\_auction](self.pendingReturns[msg.sender]) }@]
def withdraw():
    pending_amount: wei_value = self.pendingReturns[msg.sender]
    self.pendingReturns[msg.sender] = 0
    send(msg.sender, pending_amount)
\end{lstlisting}
\end{minipage}
\end{center}
\caption{Example usage of derived resources in a part of the auction contract. Ghost commands are \textcolor{red}{red} and specifications like effects-clauses (using the \code{performs} keyword) and resource declarations are \textcolor{darkgreen}{green}. Since the contract declares a resource \code{wei_in_auction} derived from wei, sending some wei to it when calling function \code{bid} will implicitly create the same amount of \code{wei_in_auction}, which then belongs to the bidder. Every bidder offers to exchange their  \code{wei_in_auction} for the auctioned \code{good} if they win the auction. When calling \code{withdraw}, previous bidders get back the wei they sent, implicitly destroying their \code{wei_in_auction}.} \label{fig:derivedexamplepart}
\end{figure}

\subsection{Further Extensions}
Our system contains a few more features that we are not able to describe fully for space reasons.
The most important is the notion of \emph{delegation}: It is sometimes useful or necessary for collaborating contracts to be able to act in each others' names when interacting with other contracts. To enable this, we allow a contract $A$ to decide to \emph{trust} another contract $B$ \wrt{} outside contract $C$, meaning that when $B$ interacts with $C$ (and only then), it can perform actions that normally only $A$ would be able to perform (\eg{} transfer $A$'s resources to someone else). As a result, all restrictions on who may execute certain ghost command that we have discussed so far are implemented modulo trust.
Since trusting someone weakens the guarantees one has for one's own resources, users must use this feature with caution; however, as with all other potentially negative effects, functions that establish new trust relations must always state that they do so in their effects-clause.

Our core methodology is easily extended in several further ways; our implementation \eg{} has support for resources with identifiers (resources whose instances can be distinguished from one another)  useful for modelling non-fungible tokens (NFTs)~\cite{ERC721}. Other generalisations are possible, \eg{} for some contracts it may be useful to have derived resources that represent ownership not of a single resource of a different type, but of different amounts of other resources. This feature (like many others) does not have to be built into the system; it can be emulated by using the existing resource model in combination with additional invariants, segment constraints etc. that represent the additional rights and constraints that would result from such resources. As we show in \secref{evaluation}, the set of features we have described gives users a sufficiently expressive model to be able to verify real contracts, while being simple enough for users to reason about.

\section{Proof Technique} \label{sec:logic}
We summarise here the formalisation of our technique as a separation logic; for space reasons, full details are relegated to \fullversionref{formalisation}.
We do so for a simple smart contract language reflecting the core of Vyper, with the following commands:\\
$
\begin{array}{rcl}
c&{::=}&\noop \mid \cassign{x}{e} \mid \cassign{\self.f}{e} \mid \ccall{x}{e}{fun}{e}{e} \mid \cseq{c}{c} \mid \cassert{e}
\end{array}
$

To reflect the design of Vyper only fields of $\self$ (the current contract) can be assigned; function calls take a second argument representing the amount of wei to send along with a call.
We assume a standard expression language with a reserved $\result$ identifier (to refer to function results in postconditions); field lookups include those on the implicit $\msg$ and $\block$ arguments. To express two-state assertions such as our segment constraints, our formalisation includes \emph{labels} $l$ denoting earlier points in execution, and expressions $\oldl{l}{e}$ denoting the value $e$ had at label $l$. We use three labels: $\preL$, representing the pre-state of the current function, $\lastL$, representing the pre-state of the current local segment, and $\callL$, representing the pre-state of the last call to another contract.

A state $\Sigma$ has the form $\langle \mathcal{H}, \mathcal{R}, \mathcal{E}, \mathcal{O}, \sigma \rangle$, where $\mathcal{H}$ is the heap (a partial map from contract addresses and fields to their values), and $\sigma$ is the current variable store. $\mathcal{E}$ is a multiset of \emph{effects} produced so far by the current function and $\mathcal{R}$ is a record containing the state of all resources declared in the current contract. In particular, for every such resource $R$, $\mathcal{R}.\mallocated{R}$ maps addresses to their balances, and $\mathcal{R}.\moffered{R}{R'}$ tracks the offers to exchange $R$ against another resource $R'$ declared in the contract. $\mathcal{R}.\mtrusted$ is a partial map from pairs of addresses to boolean values that represent whether the first address currently trusts the second; expressions can refer to these maps. Finally, $\mathcal{O}$ maps label names to pairs $(\mathcal{H}, \mathcal{R})$ that represent the heap and resource state at label $l$.

Expression evaluation in a state, denoted by $\eeval{e}{\sigma}{\mathcal{H}}{\mathcal{R}}{\mathcal{O}}$, is largely standard; most-interestingly, the evaluation of $\eeval{\oldl{l}{e}}{\sigma}{\mathcal{H}}{\mathcal{R}}{\mathcal{O}}$ is $\eeval{e}{\sigma}{\mathcal{H}'}{\mathcal{R}'}{\mathcal{O}}$, where $\mathcal{O}[l] = (\mathcal{H}', \mathcal{R}')$.

We now define our assertion language as follows:\\
$
\begin{array}{rcl}
P,Q&{::=}& \aemp \mid e \mid P \ast P \mid P \wedge P  \mid P \wand P \mid P \vee P \mid \\
& & \aperformed{E} \mid \aalloc{R}{e}{e} \mid \aoffers{R}{R}{e}{e}{e}{e}{e} \mid \atrusts{e}{e}{e}
\end{array}
$

\begin{figure}[t!]
\[\small
\begin{array}{c}
\Inf[SCall]{e_r: T}{T.\code{fun}(x)~\ensures{Q}~\performs{S}}{S' \subseteq S}
                  {\htriple{}{
\begin{array}{c}
\CInv[\oldl{last} / \old]\\
\wedge \CLC[\oldl{last} / \old] \\
\wedge e_O
\end{array}
                  }{\ccalltl{x}{e_r}{fun}{e_a}{e_v}}{
\begin{array}{c}
\aperformed{S'}[e_r/\self][x/ \result ] \\
~[\self/\msg.sender][e_a/x] \ast \\
\oldl{call}{e_O} \wedge \\
\CInv[\oldl{call} / \old] \wedge \\
\CGPost[\oldl{call} / \old] \wedge \\
Q[e_r/\self][x/ \result ] \\
~[\self/\msg.sender][e_a/x]
\wedge \\
e_N \Rightarrow \old{e_N}
\end{array}
                  }} \\
\ \\
\Inf[Frame]{\FV(R) \cap \mods(c) = \emptyset}{\htriple{}{P}{c}{Q}}.
           {R\text{ is stateless if $c$ contains a call}}
           {\htriple{}{P \ast R}{c}{Q \ast R}} \\
\ \\
\Inf[Transfer]{\htriple{}{
\begin{array}{c}
e_a \geq 0 \ast \\
(a \neq 0 \Rightarrow \\
 \atrusts{e_f}{\code{msg.sender}}{\tru}) \\
 \ast \aalloc{R}{e_f}{e_a}
\end{array}
}{\creallocate{R}{e_f}{e_t}{e_a}}{
\begin{array}{c}
\aalloc{R}{e_t}{e_a} \ast \\
(e_a \neq 0 \Rightarrow \\
 \atrusts{e_f}{\code{msg.sender}}{\tru})\\
 \ast \aperformed{\ereallocate{R}{f}{t}{a}}
\end{array}
}} \\
\end{array}
\]
\caption{Selected Hoare Logic rules; full rules included in \fullversionref{formalisation}. }\label{fig:logicshort}
\end{figure}

Assertion truth in a state is defined by a judgement $\langle \mathcal{H}, \mathcal{R}, \mathcal{E}, \mathcal{O}, \sigma \rangle \models P$ whose cases are given in \fullversionref{formalisation}.  In contrast to traditional separation logics \cite{sl}, we do not use the \emph{linear/separating} aspects of the $\ast$ and $\wand$ connectives to govern access to the (already-encapsulated) \emph{heap}, but rather for the resource state and effects concepts added by our methodology. The separating conjunction $P \ast Q$ splits the resource state and the effects into two parts; the first described by $P$ and the second by $Q$. Descriptions of constituent parts of the resource state come via assertions such as
$\aalloc{R}{e_o}{e_a}$ that prescribe that $\mathcal{R}$ is empty (no offers, no trust, and all balances are zero) \emph{except} for the balance of $e_o$, which owns exactly $e_a$ of resource $R$, and that $\mathcal{E}$ is empty (no effects).
The (multiplicative) separating conjunction builds up larger descriptions of these states; \eg{} $\aalloc{R}{e_o}{e_{a_1}} \ast \aalloc{R}{e_o}{e_{a_2}}$ is equivalent to $\aalloc{R}{e_o}{e_{a_1} + e_{a_2}}$.
Similarly, $\aperformed{E}$ states that \emph{exactly} the effects in multiset $E$ have been performed and no others (and the resource state is empty).
The interpretation of other assertions is standard for a classical separation-logic; in particular, an assertion $e$ is true only if there are \emph{no} effects in the current state.
As a result, assertions always have to describe the effects-state precisely: If a state containing $\mathcal{E}$ fulfils $P \ast \aperformed{E}$, and $P$ does not syntactically contain $\aperformed{}$-assertions, then we must have that $\mathcal{E} = E$. This is important to ensure that functions report all effects they directly cause.

To handle the various kinds of two-state specifications our methodology employs (in each of which $\old{e}$ is used to denote evaluation in the appropriate ``old'' state), we define a judgement $\fulfils{\Sigma_1}{\Sigma_2}{P}$ in which $\Sigma_1$ represents the appropriate state to use as the old one (\eg{} for local segment constraints we use the state at the start of the local segment):
\begin{definition}
For two states $\Sigma_1 = \langle \mathcal{H}_1, \mathcal{R}_1, \mathcal{E}_1, \mathcal{O}_1, \sigma_1 \rangle$ and $\Sigma_2 = \langle \mathcal{H}_2, \mathcal{R}_2, \mathcal{E}_2, \mathcal{O}_2, \sigma_2 \rangle$ we define
$\fulfils{\Sigma_1}{\Sigma_2}{P}$ if and only if $\langle \mathcal{H}_2, \mathcal{R}_2, \mathcal{E}_2, \mathcal{O}_2[last \mapsto (\mathcal{H}_1, \mathcal{R}_1)], \sigma_2 \rangle \models P[\oldl{last}{\_} / \old{\_}]$%
\end{definition}
Using this notion, we can define two-state assertion reflexivity and transitivity as follows:
\begin{definition}
An assertion $P$ is \emph{reflexive} if, for all $\Sigma_0, \Sigma_1$, if $\fulfils{\Sigma_0}{\Sigma_1}{P}$, then $\fulfils{\Sigma_1}{\Sigma_1}{P}$.
An assertion $P$ is \emph{transitive} if, for all $\Sigma_0, \Sigma_1, \Sigma_2$, if $\fulfils{\Sigma_0}{\Sigma_1}{P}$ and $\fulfils{\Sigma_1}{\Sigma_2}{P}$, then $\fulfils{\Sigma_0}{\Sigma_2}{P}$.
\end{definition}
We can now also define assertion \emph{stability}, \ie{} the fact that an assertion is preserved by another:
\begin{definition}
An assertion $P$ is \emph{stable under} $Q$, written $\stable{P}{Q}$, if,
for all $\Sigma_0, \Sigma_1, \dots, \Sigma_n$, if $\fulfils{\Sigma_0}{\Sigma_1}{P}$ and $\fulfils{\Sigma_i}{\Sigma_{i+1}}{Q}$ for all $i \in \{1, \dots, n-1\}$, then $\fulfils{\Sigma_0}{\Sigma_n}{P}$.
\end{definition}
Finally, we need to define the notion of a stateless assertion:
\begin{definition}
An assertion $P$ is \emph{stateless} if it does not refer to the current state (including resource state) or an old state \emph{except} for the pre-state (\ie{} only old-expressions with label $pre$ are allowed).
\end{definition}

We define our proof technique via a Hoare Logic formulation, whose details are given in \fullversionref{formalisation}. Three important example rules are shown here in \figref{logicshort}. Rule \rulename{SCall} is a simplified version of the most important rule: that for reasoning about calls. Here, ordinary ($\CLC$) and transitive segment constraints ($\CInv$) must hold at the end of every local segment; we therefore require them to be true in the precondition of the Hoare triple, using the state labelled ``last'' as old state. This notion of ``last'' state is reset in the postcondition to the new current state (we begin a new local segment), as indicated by $e_N \Rightarrow \old{e_N}$ which can be instantiated for any $e_N$. A similar connection can be made to facts known before the call via $e_O$.

The frame rule $\rulename{Frame}$ is non-standard in that it ensures that \emph{no} information about the last state or the current state can be framed around calls; this represents the fact that the entire contract state can change with every call, due to unknown transitive calls.
After a call, one may nonetheless assume the transitive segment constraints and function constraints \wrt{} the call's pre-state. To remember information about said pre-state, we use the same trick as before, and allow assuming any expression $e_O$ after a call about its pre-state that was known to be true before the call.

Each ghost command of \secref{resources} gets a corresponding Hoare Logic rule (\eg{} \rulename{Transfer} shown here), which: (a) checks that the participant for whom the command has a negative effect (\eg{} giving away resource) trusts the current $\code{msg.sender}$ (typically, this is simply who they are), (b) checks that required resources for the command are available, consuming them, (c) adds appropriate new resources in the postcondition assertion, and (d) records the effect that was performed.

The rule for constructors (not shown here) performs the necessary checks of transitivity and reflexivity of transitive segment and function constraints, and ensures that transitive segment constraints fulfil stability criteria described previously. %

\section{Implementation and Evaluation} \label{sec:implementation}\label{sec:evaluation}
We have implemented our work in \twovyper, an automated verification tool for the Vyper language. \twovyper is open source\footnote{\url{https://github.com/viperproject/2vyper}}; it reuses the standard Vyper compiler to type-check input programs. It encodes Vyper programs and specifications into the Viper intermediate verification language~\cite{viper}, and uses Viper's infrastructure and ultimately the SMT-solver Z3~\cite{z3} to verify the program or otherwise return errors and counterexamples.

While less commonly used than Solidity, Vyper puts a stronger focus on correctness and simplicity, by preventing some errors on the language level (such as over- or underflows, which automatically revert the transaction, unlike in Solidity) and omitting some language features that make code more difficult to reason about (such as inheritance). \twovyper supports the entire current Vyper language (and several previous versions) and is intended for full-fledged verification of real-world contracts.

\begin{figure}
\begin{center}
\begin{minipage}[t]{0.8\textwidth}
\begin{lstlisting}
contract Client:
  def client():
    self.token.approveAndCall(self.service, amount, data

contract Token:
  def transferFrom(from : address, amount: uint256):
    # transfer 'amount' from 'from' to msg.sender
    # if msg.sender has a sufficient allowance

  #@ performs: revoke[token <-> token](1, 0, to=spender)
  #@ performs: offer[token <-> token](1, 0, to=spender, times=amount)
  def approveAndCall(spender: address, amount: uint256, data: bytes[1024]):
    #@ revoke[token <-> token](1, 0, to=spender)
    self.allowances[msg.sender][spender] = amount
    #@ offer[token <-> token](1, 0, to=spender, times=amount)
    ERC1363Spender(spender).onApprovalReceived(msg.sender, amount, data)

contract Service implements ERC1363Spender:

  def onApprovalReceived(sender: address, amount: uint256, data: bytes[1024]):
    self.token.transferFrom(sender, amount)
    self.performService(sender, amount, data)
\end{lstlisting}
\end{minipage}
\end{center}
\caption{Simplified code example showing the functionality of ERC1363 payments. Function \code{approveAndCall} also shows the specification syntax used by \twovyper.
The client calls \code{approveAndCall} on the token contract and supplies as arguments both the service provider and the input for the requested service. The token contract stores that the service provider may take tokens from the client (in field \code{allowances}), and then invokes \code{onApprovalReceived} on the service provider, which re-entrantly calls \code{transferFrom} to take its tokens and then performs the service.
This architecture intentionally uses re-entrancy to allow clients to do in one transaction what would usually require two (one for setting the allowance, one for invoking the service).
}\label{fig:erc1363}
\end{figure}

\twovyper specifications are written as $\sharp@$ comments in the source code, and use Vyper syntax wherever possible. Fig.~\ref{fig:erc1363} shows a simplified excerpt of a function annotated with specifications and containing ghost commands for resource manipulation. %
In addition to the core correctness properties we have focused on in this paper, \twovyper also supports reasoning about additional language features (\eg{} events) and has additional specification constructs to prove specific kinds of liveness properties (\eg{} that auction bidders can eventually get their Ether back; it is not stuck in the auction contract forever) that can be converted to safety properties for verification.

\subsection{Evaluation Examples}
We have evaluated our approach on a number of real-world smart contracts focusing on existing contracts written in Vyper as well as those involving pertinent features such as inter-contract collaboration or re-entrancy bugs~\cite{impl-vyper-examples,impl-dao,impl-1363,impl-verx,impl-solidity-examples,impl-serenus,impl-uniswap}.
We manually translated some examples without Vyper implementations from Solidity to Vyper.

\begin{table}[t]
\small
\begin{tabular}{l|l|r|r|r|r|r}
Application                  & Contract             & LOC & Ann. & IF LOC & IF Ann. & $T$  \\ \hline
auction              & auction              & 63     &  30   &  -      &    -    & 12.72  \\
auction\_token       & auction\_token$^{\dagger\dagger}$      & 96    &  37   & 88       &  33      & 23.67  \\
DAO                  & DAO*                 &  17   &  2   &   -     &   -     &   5.13 \\
ERC20               & ERC20                & 98    &  31   &  67      &   25     & 11.51  \\
ERC721               & ERC721               & 178    & 32    &    -    &   -     &  15.95 \\
ERC1363             & ERC1363              & 142    & 31    &  88       &  33  &  22.59 \\
ICO                  & gv\_option\_token      & 98    & 26    &  86      &  36      &  10.03 \\
                     & gv\_token             &  121    & 24    &  107      &   49     &  15.21 \\
                     & gv\_option\_program*   &  86   &  12   &   154     &  67      &  29.19 \\
                     & ico\_alloc*           & 159    & 30    &  261      &  116      &  101.86 \\
Mana                 & token*                &  18   &  3   &  50     &   25    &  2.78 \\
                     & crowdsale*            & 42    &  14   &   50     &  25      &  5.77 \\
                     & continuous\_sale*      &  36   & 8    &  50      &   25     &  3.88 \\
VerX\_overview        & escrow               &  60   &  11   &    65    &   33     &  6.36 \\
                     & crowdsale            &  41   &   9  &   65     &   33     &  6.35 \\
safe\_remote\_purchase & safe\_remote\_purchase & 71    &  29   &  -      &  -      &  16.84 \\
serenuscoin          & serenuscoin          & 103    &  4   &   -     &  -      &  6.40 \\
Uniswap V1              & Uniswap$^{\dagger}$            & 398    &  115    &  105      &  45      & 112.81
\end{tabular}
\caption{Evaluated examples. LOC are total lines of code,  \emph{including} specification, excluding comments and whitespace. Ann. are lines of specification. IF LOC and IF Ann. have the same meaning for the interfaces that were required to verify the contract, and $T$ is verification time in seconds. Contracts marked with a star are simplified versions of the original; applications marked with one or two daggers collaborate with an external ERC20 or ERC1363 contract, respectively (accessed through an interface).}\label{tab:examples}
\end{table}

Table~\ref{tab:examples} shows the examples; while many consist of a single contract, several either consist of multiple collaborating contracts or of a single contract interacting with external contracts via interfaces. We include several examples of complex code used in practice, \eg{} ERC tokens, the first version of the Uniswap contract (the largest decentralized exchange and fourth-largest cryptocurrency exchange overall), and an application used to implement the Genesis Vision ICO, which raised 2.8 million USD in 2017. Most contracts were verified in their entirety; for the ICO, we made some small simplifications (in particular, we cut out two option tokens that behaved exactly like a third one and so added nothing of interest); for the Mana and DAO contracts, we focused on specific parts demonstrating inter-contract invariants and a re-entrancy bug, respectively.

\subsection{Verified Properties}
We now describe the functionality, verified properties, and used specification constructs for some of our examples; if no specific properties are stated, we verified a full functional specification. \fullversionref{other-examples} contains descriptions of the remaining examples.

\myparagraph{ERC20, auction and auction\_token:}
We have verified an extended version of the auction contract from Fig.~\ref{fig:exampleauction} and proved all properties mentioned throughout this paper.
We have also verified the widely-used standard Vyper ERC20 implementation, which is a more complex version of the token contract in Fig.~\ref{fig:exampletoken}, by declaring a token resource and annotating all functions with the resource operations they perform. We also used segment constraints to specify when the contract triggers \emph{events}, which are a means for the contract to log which transactions have happened in a way that is visible outside the blockchain, and which can easily be specified using segment constraints.

Finally, we have verified a variant of the auction that deals in custom tokens instead of wei against an ERC1363 interface \cite{ERC1363} (see below) annotated with resource-based specifications.

\myparagraph{DAO:}
We extracted the buggy part of the DAO contract that led to the loss of ca. 50 million USD~\cite{dao}.
Our implementation declares a derived resource for Ether by default (\ie{} it assumes that Ether sent to a contract should still conceptually belong to the sender unless otherwise specified).
As a result, when the contract tries to send Ether to an address, an error is reported by default, since our resource model requires the user to justify this action by showing that Ether is only sent to its rightful owner. Since this is not always the case, the contract will be rejected.

\myparagraph{ICO:}
We verified four contracts that implement the Initial Coin Offering (ICO) for Genesis Vision. The ICO progresses in stages, first selling options, then starting the ICO for option holders, and subsequently for the public. Verification required all specification constructs we have presented, \eg{} function constraints to describe guarantees made by locks, transitive segment constraints to preserve information across calls (\eg{} that the main token, once unfrozen, will never be frozen again), and resource specifications modeling the option token and main token. We used trust to allow that an administrator can freely access other's tokens, which our technique normally rules out.
Importantly, we required proving multiple inter-contract invariants to coordinate the four contracts that implement the ICO, \eg{} to prove that the main token will be frozen in its first stages.

Some (inter-contract) properties of this example have also been verified in VerX~\cite{verx}. Notably however, VerX requires the code of all involved contracts at once and does not allow using interface abstractions. In contrast, we use interfaces annotated with specifications to verify each contract modularly. Additionally, while we prove every property proved by VerX, we also proved additional properties (\eg{} all standard resource properties such as non-duplicability and ownership, and that the resource operations the contract performs are the expected ones).

\myparagraph{Uniswap V1:}
Uniswap is a popular application that consists of many different exchanges, which together allow clients to exchange different tokens against each other, using Ether as an intermediary.
A single exchange is responsible for a single token and, if it wants to buy other tokens, contacts the respective exchange contracts for those other tokens. We declared the desired resource-effects for each function and proved the exchange contract correct \wrt{} them. Again, we did so modularly, using a standard ERC20 interface for its token contract and another interface for other exchanges.

\myparagraph{VerX overview:}
We verified the crowdsale application (consisting of two contracts) from the VerX paper, which again included an inter-contract invariant that we verified \emph{modularly} using interfaces and privacy constraints. Additionally, since one of the involved contracts implements a state machine, we used transitive segment constraints to define valid transitions between states (\eg{} once the contract is in the ``refund'' state, it remains in this state).

\myparagraph{ERC1363:}
ERC1363 is a new token standard \cite{ERC1363} that combines into one transaction what ERC20 does in several. Fig.~\ref{fig:erc1363} illustrates how this contract intentionally uses re-entrancy in a way that is not ECF and thus cannot be verified using approaches such as VerX.

\myparagraph{Conclusion:} Our evaluation shows that our specification constructs allow specifying and verifying a wide variety of different properties for real-world contracts. In particular, we can modularly prove inter-contract properties, we can model the resources and resource transactions of different, complex contracts using our resource system (and find typical errors by default), and we can give guarantees for functional correctness  and access control even in the presence of unbounded re-entrancy, which allows us to support contracts that employ re-entrancy by design.

\subsection{Performance and Annotation Overhead}
Table~\ref{tab:examples} shows the total lines of code of each contract (excluding comments and whitespace, including specification) as well as the lines of annotations we require, and the lines of code and specifications of all interfaces required to verify each contract, as well as the verification time required by \twovyper.
Times were measured by averaging over ten runs, running on a warmed-up JVM.

On our test system (a 12-core Ryzen 3900X with 32GB RAM running Ubuntu 20.04),
most contracts can be automatically verified in 5-25 seconds; the two contracts with the longest verification time, both of which are from complex real-world applications, take between 1.5 and 2 minutes. Considering the strong guarantees afforded by our methodology and tool, we believe even the longest of these times is quite acceptable in practice.

The number of lines required for specifications is less than the number of lines of actual code for every contract.
This comparatively modest specification overhead is partly due to our domain-specific resource specifications that allow users to express complex properties in a concise way, and partly due to the design of Vyper, which simplifies verification.
Overall, considering the potential financial losses resulting from incorrect smart contracts, writing this amount of specification in exchange for strong functional correctness guarantees is clearly worthwhile.

In conclusion, our technique enables concisely specifying complex correctness properties of (collaborating) contracts, while allowing for modular verification that can be automated efficiently.

\section{Related work} \label{sec:related}

A lot of recent work has focused both on finding problems in smart contracts and on proving their absence. Atzei et al.~\cite{DBLP:conf/post/AtzeiBC17} and Luu et al.~\cite{DBLP:conf/ccs/LuuCOSH16} each list different kinds of potential attacks and problems specific to smart contracts.
A number of tools have been built to automatically find such problems (\eg{} resulting from re-entrancy) using either syntactic patterns~\cite{DBLP:conf/ccs/TsankovDDGBV18,DBLP:conf/icse/TikhomirovVITMA18,DBLP:conf/iccsp/LaiL20}, bounded symbolic execution~\cite{DBLP:conf/isola/AltR18,DBLP:journals/corr/abs-1907-03890,DBLP:conf/ccs/LuuCOSH16,DBLP:conf/acsac/NikolicKSSH18} or data flow analyses~\cite{DBLP:conf/icse/FeistGG19}. However, most of these tools are not designed to be sound and can therefore miss errors in real contracts~\cite{DBLP:conf/icse/TikhomirovVITMA18,DBLP:conf/icse/FeistGG19,DBLP:conf/ccs/TsankovDDGBV18}. Additionally, none of these tools allow proving custom functional properties.

Recent work has studied the difference between harmless re-entrant executions and re-entrancy vulnerabilities~\cite{DBLP:conf/setta/CaoW20}. In particular, \citet{DBLP:journals/pacmpl/GrossmanAGMRSZ18} have introduced the notion of effectively callback free objects, for which (in a smart contract setting) re-entrancy does not introduce any behaviours that are not present in executions without re-entrancy. They provide an algorithm for dynamically checking for ECF-violations and study the decidability of statically proving that a contract is ECF. More recently, Albert et al.~\cite{DBLP:journals/pacmpl/AlbertGRRRS20} show a static analysis for deciding ECF based on commutativity and projection.

A number of tools aim to achieve verification of custom functional properties for Ethereum contracts, either on the level of the Solidity language~\cite{DBLP:journals/corr/abs-1907-04262,DBLP:conf/ndss/KalraGDS18} or on the level of EVM bytecode~\cite{DBLP:conf/csfw/HildenbrandtSRZ18} - to the best of our knowledge, \twovyper is the first verifier specifically aimed at the Vyper language. EVM-based verification is not specific to any high-level source language and does not rely on the correctness of the compiler; however, specifications tend to become much more complex on the EVM-level, where high-level abstractions of the source language are lost.
Verification tools are either based on SMT solvers~\cite{DBLP:journals/corr/abs-1907-04262,verx}, model checking with code generation~\cite{DBLP:conf/fc/MavridouLSD19}, matching logic~\cite{DBLP:conf/csfw/HildenbrandtSRZ18}, CHC solving~\cite{DBLP:conf/ndss/KalraGDS18}, or interactive theorem provers~\cite{DBLP:conf/fc/Hirai17}, which offer different levels of automation and expressiveness.
Existing verification tools that offer dedicated, higher level specification languages (\eg{} \citet{DBLP:journals/corr/abs-1907-04262}) typically support single-state contract invariants,
but offer no special support for reasoning in the presence of arbitrary re-entrancy beyond that, resulting in imprecision.
VerX~\cite{verx} and VeriSolid~\cite{DBLP:conf/fc/MavridouLSD19} can express temporal properties, which subsume ordinary history constraints; however,
VerX explicitly only targets contracts that are ECF, and VeriSolid prevents all re-entrancy by generating code that uses locks throughout.
Addionally, none of the existing Ethereum verifiers support resource-based specifications.

To the best of our knowledge, the only tools capable of proving user-defined inter-contract properties are VerX and VeriSolid.
The former requires the source code of all involved contracts and is therefore not contract-modular, unlike our approach.
The latter uses model checking to prove temporal logic properties on a higher-level model of the contracts and their allowed interactions, and generates Solidity code from this model which is correct-by-design~\cite{DBLP:conf/icbc2/NelaturuMVL20}. In contrast to our approach, VeriSolid does not allow reasoning directly about the code of an existing contract (though contracts can be imported into the model).

Researchers have proposed a number of new smart contract languages that aim to simplify verification and/or make it more difficult to write incorrect code~\cite{move,DBLP:journals/pacmpl/SergeyNJ0TH19,DBLP:conf/icse/Coblenz17}. In particular, the Move language~\cite{move} offers resources on the programming language level. Unlike our resource model, these resources are stateful (in fact, \emph{all} state in Move is stored in resources) and do not have a one-to-one correspondence to physical goods or currency: Receiving $n$ coins, for example, is implemented in Move by adding $n$ to the value stored in one's existing single coin resource, since every address can have at most one resource of every kind. %
While a linear type system ensures that resources are not duplicated in third party code, the module that defines a resource may modify resource state in arbitrary ways.
As a result, incorrect module implementations in Move can potentially violate the properties guaranteed by our resource system (\eg{} that resources cannot be taken away from their owners); on the other hand, Move's system allows users to manually implement more complex resource models than ours. Finally, an SMT-based verifier for custom properties of Move programs exists~\cite{DBLP:conf/cav/ZhongCQGBPZBD20} but currently does not offer special support for specifying resource transfers.

To the best of our knowledge, there are three existing approaches for reasoning about (object-oriented) programs in the presence of unverified code. First, \citet{DBLP:conf/fase/Drossopoulou0ME20} have introduced \emph{holistic} specifications, which (unlike traditional ones) express \emph{necessary} conditions for an effect to happen, in a setting with arbitrary re-entrancy. They can express \eg{} that if a user's token balance decreases, then they either asked to transfer tokens themselves, or another user with a sufficient allowance must have done so.
While this kind of property is similar to ones ensured by our resource system, it is not built-in and must be specified manually.
Additionally, holistic specifications do not provide support for reasoning about the post-state of calls with arbitrary re-entrancy, and the required (non-standard) reasoning has not been automated, whereas the proof obligations generated by our approach can be checked and automated using standard techniques.

Second, software architectures based on object capabilities~\cite{objectcapabilities} and object capability patterns~\cite{capabilitypatterns} can be used to encapsulate object state so that properties can be maintained even in an unverified environment. The central idea of object capabilities is to withhold the reference to an encapsulated object from unauthorized third parties, and thereby control who may invoke operations on the object. It is therefore crucial that third parties cannot forge capabilities and thereby obtain unintended access to the encapsulated object.
However, since this is not the case in typical smart contract languages
(contract addresses are not opaque and can be obtained in various ways, not only by receiving them as an intended capability from another contract),
the conditions required for capability-based reasoning are not satisfied in this setting.

Third, \citet{DBLP:conf/popl/Agten0P15} apply separation logic in a context with unverified code by using runtime checks at the boundary between verified and unverified code to ensure that
the unverified code has not modified memory it was not permitted to modify.
In contrast, our work relies on language encapsulation to ensure this property and therefore does not require runtime checks, which are especially undesirable in a smart contract setting due to the associated gas cost.

\section{Conclusion} \label{sec:conclusion}
In this paper, we have presented a novel approach for specification and verification of Ethereum smart contracts. Our methodology exploits the features of Ethereum, such as strong encapsulation, to provide guarantees even in the presence of arbitrary re-entrancy, and provides domain-specific specification constructs for resources that make specification both more intuitive and less error-prone. Our evaluation shows that out methology can be implemented efficiently and is capable of expressing and proving complex functional specifications for real-world contracts.

\begin{acks}
We gratefully acknowledge support from Swiss Stake GmbH.
\end{acks}

\bibliographystyle{ACM-Reference-Format}
\bibliography{papernocomments}

\iffullversion

\appendix

\section{Resource ghost commands and effects}\label{app:ghost-commands}

In this section, we give a more detailed description of the ghost commands that can be used to modify the resource state, and the effects they cause.

\subsubsection*{Creating and Destroying Resources.}
Initially, every address has a balance of zero for every resource.
Resource can be created by executing the ghost command
$\ccreate{R}{e_c}{e_t}{e_a}$, which means that $e_c$ creates $e_a$ amount of resource $R$ and allocates it to $e_t$, and which has the effect $\ecreate{R}{e_t}{e_a}$.
Since this command states that $e_c$ is the one creating the resource, this command may be performed only if $e_c$ is the current $\msg.\code{sender}$, \ie{} the caller of the function that contains this ghost command (modulo trust).
Creating a resource additionally requires owning the \emph{right} to do so, which we represent by a special resource $\rcreator{R}$.
In the constructor of the contract, the caller of the constructor automatically gets the right to create such creator resources (\ie{} they get a resource $\rcreator{\rcreator{R}}$ for every resource type $R$ declared by the contract) and can then set up the contract so that the parties that are intended to have minting rights each own a creator resource at the end of the constructor. %

In function \code{mint} of the token contract, which increases the number of tokens in the contract state, we would have to insert the command $\ccreate{R}{\code{self.minter}}{\code{to}}{\code{amount}}$ to preserve the relation between $\mallocated{\mathit{token}}$ and the contract state. Verifying \code{mint} then requires showing that the caller is  \code{self.minter}, which we know because of the assertion in the first line, and that \code{self.minter} owns a creator resource for \text{token}. To show the latter, we must create such a resource in the contract's constructor (not shown in Fig.~\ref{fig:exampletoken}), and record the fact that the minter owns it in an additional invariant.

Conversely, when executing the command $\cdestroy{R}{e_f}{e_a}$, $e_f$ destroys $e_a$ amount of its reserves of resource $R$, which requires that $e_f$ actually has that amount of resource, and, as before, that $e_f$ is (or trusts) the $\msg.\code{sender}$. This ghost command has the effect $\edestroy{R}{e_f}{e_a}$.

\subsubsection*{Transfers.}

The ghost command $\creallocate{R}{e_f}{e_t}{e_a}$ transfers $e_a$ amount of resource $R$ from $e_f$ to $e_t$, and requires that $e_f$ owns that amount and is (or trusts) the $\msg.\code{sender}$.
This last requirement, along with the similar requirement for destroying resources, is what guarantees that once a contract owns a resource, no other contract can take it away or destroy it. The command has the effect $\ereallocate{R}{e_f}{e_t}{e_a}$.

\subsubsection*{Offers and Exchanges.}

The resource transfers just shown can only happen directly at the command of the sender, at the moment the sender requests them. In practice, this is not always sufficient to model different kinds of resources. Therefore, as we explain in the main body of the paper, we allow two exceptions from this general requirement; the first is that contracts can \emph{offer} to perform specific resource \emph{exchanges} with other contracts at some later point in time. If both parties have offered an exchange, the exchange can then be performed at any point, without requiring further agreement from the involved parties.

The command $\coffer{R_1}{R_2}{e_f}{e_t}{e_{a_1}}{e_{a_2}}{e_n}$ creates $e_n$ offers from $e_f$ to contract $e_t$ to exchange $e_{a_1}$ amount of $R_1$ against $e_{a_2}$ amount of $R_2$.
The command $\crevoke{R_1}{R_2}{e_f}{e_t}{e_{a_1}}{e_{a_2}}{e_n}$ revokes $e_n$ offers from $e_f$ to contract $e_t$ to exchange $e_{a_1}$ amount of $R_1$ against $e_{a_2}$ amount of $R_2$.
Both commands can only be executed by $e_f$ or someone trusted by $e_f$. However, they do \emph{not} require that $e_f$ actually \emph{owns} these amounts of the specified resources: contracts can offer exchanges that they could not actually perform at the time of making the offer, which can then potentially be performed at a later point once they have the necessary resources. Like the previous ghost commands, both of these ghost commands have corresponding effects with the same form.

The command $\cexchange{R_1}{R_2}{e_f}{e_t}{e_{a_1}}{e_{a_2}}$ performs an exchange between $e_{a_1}$ amount of $R_1$ from $e_f$ and $e_{a_2}$ amount of $R_2$ from $e_t$. Executing it requires that both $e_f$ and $e_t$ have made an offer to perform such an exchange, and consumes the offer. There is one exception to this: if either $e_{a_1}$ or $e_{a_2}$ are zero, \ie{} the exchange simply gives a resource to one party without requiring anything in return, no offer is required from the party receiving the resource.

Unlike any other ghost command, there are no requirements as to who can execute this command (\ie{} who the $\msg.\code{sender}$ is), since its effect is one that all affected parties have already agreed to previously. The effect $\cexchange{R_1}{R_2}{e_f}{e_t}{e_{a_1}}{e_{a_2}}{n}$ states that $n$ such exchanges have happened.

Note that, once a contract makes an offer to exchange some amount of a resource, it is no longer guaranteed that no other contract can take that resource away from it; instead, the resulting guarantee is that if some other contract takes resources away, this happens only in the form of an exchange the initial contract has previously agreed to.

\subsubsection*{Trust and Delegation.}
The second exception that allows resources to be managed by someone that is not the owner is, as we briefly hint at in the main body of the paper, that
a contract can \emph{trust} another contract to act (\eg{} to transfer resources or make offers) in its place.
One common use case for this is that multiple contracts collaborate and represent the same party, and need to be able to act in each other's name.
Some token contract standards, \eg{} ERC721~\cite{ERC721}, have a built-in mechanism to enable exactly this pattern.

Trust is \emph{local}, \ie{} $A$ can trust $B$ to act in its name \emph{only} when interacting with contract $C$.
Like for offers and balances, every contract has a ghost map $\mtrusted$ that represents which contracts trust which other contracts when interacting with the current contract.
For all commands we have previously discussed, the requirement is therefore not actually that the $\msg.\code{sender}$ \emph{is} the respective source address/creator/..., but that the $\msg.\code{sender}$ is \emph{trusted} by the source address/... (since every contract implicitly trusts itself).

Trust can be modified via the ghost command $\ctrust{e_c}{e_v}$, which sets the current caller's trust to the contract at address $e_c$ (when interacting the current contract) to the boolean value $e_v$. The effect $\etrust{e_{c_1}}{e_{c_2}}{e_v}{e_{\mathit{ctr}}}$ states that $e_{c_1}$ has set its trust to the contract at address $e_{c_2}$ when interacting with contract $e_{\mathit{ctr}}$ to $e_v$; as a result, the ghost command  $\ctrust{e_c}{e_v}$ causes the effect $\etrust{\code{msg.sender}}{e_{c}}{e_v}{\self}$.

Note that the ability to extend trust cannot be delegated, \ie{} $\ctrust{e_c}{e_v}$ always modifies the trust of the $\msg.\code{sender}$, not of anyone else.

\section{Definitions and Proof Rules}\label{app:formalisation}

\begin{figure}
$\small
\begin{array}{rcl}
\langle \mathcal{H}, \mathcal{R}, \mathcal{E}, \mathcal{O}, \sigma \rangle \models \aemp &\Leftrightarrow& \mathcal{R} = \emptyr  \wedge \mathcal{E} = \emptymultiset \\
\langle \mathcal{H}, \mathcal{R}, \mathcal{E}, \mathcal{O}, \sigma \rangle \models e &\Leftrightarrow& \eeval{e}{\sigma}{\mathcal{H}}{\mathcal{R}}{\mathcal{O}} \wedge \mathcal{E} = \emptymultiset \\
\langle \mathcal{H}, \mathcal{R}, \mathcal{E}, \mathcal{O}, \sigma \rangle \models P \ast Q &\Leftrightarrow& \exists \mathcal{R}_1, \mathcal{R}_2, \mathcal{E}_1, \mathcal{E}_2 \ldotp \mathcal{R} = \mathcal{R}_1 \uplus  \mathcal{R}_2 \wedge \mathcal{E} = \mathcal{E}_1 \mcup  \mathcal{E}_2 \wedge \\
& &  \langle \mathcal{H}, \mathcal{R}_1, \mathcal{E}_1, \mathcal{O}, \sigma \rangle \models P \wedge \langle \mathcal{H}, \mathcal{R}_2, \mathcal{E}_2, \mathcal{O}, \sigma \rangle \models Q \\
\langle \mathcal{H}, \mathcal{R}, \mathcal{E}, \mathcal{O}, \sigma \rangle \models P \wedge Q &\Leftrightarrow&  \langle \mathcal{H}, \mathcal{R}, \mathcal{E}, \mathcal{O}, \sigma \rangle \models P \wedge \langle \mathcal{H}, \mathcal{R}, \mathcal{E}, \mathcal{O}, \sigma \rangle \models Q \\
\langle \mathcal{H}, \mathcal{R}, \mathcal{E}, \mathcal{O}, \sigma \rangle \models P \wand Q &\Leftrightarrow&  \forall \mathcal{R}', \mathcal{E}' \ldotp \langle \mathcal{H}, \mathcal{R}', \mathcal{E}', \mathcal{O}, \sigma \rangle \models P \Rightarrow \\
&& \langle \mathcal{H}, \mathcal{R} \uplus \mathcal{R}', \mathcal{E} \mcup \mathcal{E}', \mathcal{O}, \sigma \rangle \models Q  \\
\langle \mathcal{H}, \mathcal{R}, \mathcal{E}, \mathcal{O}, \sigma \rangle \models P \vee Q &\Leftrightarrow&  \langle \mathcal{H}, \mathcal{R}, \mathcal{E}, \mathcal{O}, \sigma \rangle \models P \vee \langle \mathcal{H}, \mathcal{R}, \mathcal{E}, \mathcal{O}, \sigma \rangle \models Q \\
\langle \mathcal{H}, \mathcal{R}, \mathcal{E}, \mathcal{O}, \sigma \rangle \models \aperformed{E} &\Leftrightarrow&  \mathcal{R} = \emptyr \wedge \mathcal{E} = E  \\
\langle \mathcal{H}, \mathcal{R}, \mathcal{E}, \mathcal{O}, \sigma \rangle \models \aalloc{R}{e_o}{e_a} &\Leftrightarrow&  \mathcal{R} = \emptyr[\mallocated{R} := b] \wedge \mathcal{E} = \emptymultiset \\
 \text{where} && b= [o \mapsto a, \_ \mapsto 0],   \\
&& o = \eeval{e_o}{\sigma}{\mathcal{H}}{\mathcal{R}}{\mathcal{O}}, a = \eeval{e_a}{\sigma}{\mathcal{H}}{\mathcal{R}}{\mathcal{O}} \\
\langle \mathcal{H}, \mathcal{R}, \mathcal{E}, \mathcal{O}, \sigma \rangle \models \aoffers{R_1}{R_2}{e_f}{e_t}{e_{a_1}}{e_{a_2}}{e_n} &\Leftrightarrow&  \mathcal{R} = \emptyr[\moffered{R_1}{R_2} := o]  \wedge \mathcal{E} = \emptymultiset  \\
\text{where} && o = [(f, t, a_1, a_2) \mapsto n, \_ \mapsto 0], \\
&& f = \eeval{e_f}{\sigma}{\mathcal{H}}{\mathcal{R}}{\mathcal{O}}, t = \eeval{e_t}{\sigma}{\mathcal{H}}{\mathcal{R}}{\mathcal{O}} \\
&& a_1 = \eeval{e_{a_1}}{\sigma}{\mathcal{H}}{\mathcal{R}}{\mathcal{O}}, a_2 = \eeval{e_{a_2}}{\sigma}{\mathcal{H}}{\mathcal{R}}{\mathcal{O}},\\
&& n = \eeval{e_n}{\sigma}{\mathcal{H}}{\mathcal{R}}{\mathcal{O}} \\
\langle \mathcal{H}, \mathcal{R}, \mathcal{E}, \mathcal{O}, \sigma \rangle \models \atrusts{e_{c_1}}{e_{c_2}}{e_v} &\Leftrightarrow&  \mathcal{R} = \emptyr[\mtrusted := t]  \wedge \mathcal{E} = \emptymultiset \\
\text{where} && t = [(c_1, c_2) \mapsto v], v = \eeval{e_v}{\sigma}{\mathcal{H}}{\mathcal{R}}{\mathcal{O}}, \\
&& c_1 = \eeval{e_{c_1}}{\sigma}{\mathcal{H}}{\mathcal{R}}{\mathcal{O}}, c_2 = \eeval{e_{c_2}}{\sigma}{\mathcal{H}}{\mathcal{R}}{\mathcal{O}}
\end{array}
$
\caption{Definition of assertion truth in a state. If $r$ is a record, $r[f := v]$ updates field $f$ of the record to value $v$. Operator $\uplus$ for resource states is defined s.t. it performs pointwise addition for balance and offer maps, and combination of partial functions with disjoint domains for the $\mtrusted$ map. $\emptyr$ denotes an empty resource state (\ie{} all balances are zero, there are no offers, and the domain of the partial trust map is empty). We write $\emptymultiset$ for the empty multiset and $\mcup$ for multiset union.} \label{fig:assertion-validity}
\end{figure}

Assertion truth in a state is defined by a judgement $\langle \mathcal{H}, \mathcal{R}, \mathcal{E}, \mathcal{O}, \sigma \rangle \models P$ whose cases are given in Fig.~\ref{fig:assertion-validity}.  As commented in \secref{logic}, in contrast to traditional separation logics \cite{sl}, we do not use the \emph{linear/separating} aspects of the $\ast$ and $\wand$ connectives to govern access to the (already-encapsulated) \emph{heap}, but rather for the resource state and effects concepts added by our methodology. The separating conjunction $P \ast Q$ splits the resource state and the effects into two parts; the first described by $P$ and the second by $Q$. Descriptions of constituent parts of the resource state come via assertions such as
$\aalloc{R}{e_o}{e_a}$ that prescribe that $\mathcal{R}$ is empty (no offers, no trust, and all balances are zero) \emph{except} for the balance of $e_o$, which owns exactly $e_a$ of resource $R$, and that $\mathcal{E}$ is empty (no effects).
The (multiplicative) separating conjunction builds up larger descriptions of these states; \eg{} $\aalloc{R}{e_o}{e_{a_1}} \ast \aalloc{R}{e_o}{e_{a_2}}$ is equivalent to $\aalloc{R}{e_o}{e_{a_1} + e_{a_2}}$.
The assertions $\aoffers{R_1}{R_2}{e_f}{e_t}{e_{a_1}}{e_{a_2}}{e_n}$ and $\atrusts{e_{c_1}}{e_{c_2}}{e_v}$ form the analogous base cases prescribing offers and trust. Note that $\aalloc{R}{e_o}{e_a}$ can be used alongside assertions containing references to the $\mallocated{R}$ map; both are useful in different contexts. For example, an invariant that states that the allocation of $token$ in the resource state is stored in the contract in the field $\code{self.tokens}$ can be easily expressed as $\mallocated{token} = \code{self.tokens}$, whereas the proof rules for framing and resource commands, which we will show later, are much easier to express using the exact assertion $\aalloc{R}{e_o}{e_a}$.

We write $\ContractState{a}$ to refer to the entire state, including the resource state, of the contract at address $a$.
We denote (the conjunction of) the primary contract's transitive segment constraints (which may refer to other contracts' states) by $\CInv$, its segment constraints by $\CLC$, and its function constraints by $\CGPost$. The latter two may only refer to the primary contract's state.
By $\GInv$, we denote the conjunction of the transitive segment constraints of all known interfaces.
By $\CCallerPrivate{e_1}{e_2}$, we denote the conjunction of the reflexive and transitive assertions $P$ from the privacy constraints of all known interfaces \emph{except} those in set $e_2$, instantiated for $e_1$.
In all those specification constructs, the old state is referred to without a label (simply as $\old{e}$), since the kind of specification construct determines which old state it refers to.

The proof rules are shown in Fig.~\ref{fig:logic1} and Fig.~\ref{fig:logic2}.
$\FV (P)$ are the free variables in $P$,  $\mods(c)$ are the variables modified by $c$. We write $e[e_1/e_2]$ to substitute all occurrences of $e_2$ in $e$ by $e_1$.

\begin{figure}[ht!]
\[\small
\begin{array}{c}
\Inf[Assert]{\htriple{}{\tru}{\cassert{e}}{e}}
\quad \quad
\Inf[Assign]{\htriple{}{Q[e/x]}{\cassign{x}{e}}{Q}} \\
\ \\
\Inf[Write]{\htriple{}{Q[(e' = e_1 ? e_2 : e'.f)/e'.f]}{\cassign{e_1.f}{e_2}}{Q}} \\
\ \\
\Inf[Call]{\CInv' = \CInv[\oldl{last}{\_} / \old{\_}]}.
          {~[(e' = \self ? (\code{self.balance} - e_a) : e'\code{.balance}) / e'\code{.balance}]}.
          {\CLC' = \CLC[\oldl{last}{\_} / \old{\_}][(e' = \self ? (\code{self.balance} - e_a) : e'\code{.balance}) / e'\code{.balance}]}.
          {e_r: T}{T.\code{fun}(x)~\ensures{Q}~\performs{S}}{S' \subseteq S''}.
          {S'' = S[e_r/\self][x/ \result ][\self/\msg.sender][e_a/x]}.
                  {\secondary{e_r} \Rightarrow \stable{\CInv \wedge e_S}{\GInv \wedge \CCallerPrivate{\self}{\{e_r\}} \wedge \ContractState{\self} = \old{\ContractState{\self}}}}.
                  {\secondary{e_r} \Rightarrow \stable{\CInv \wedge e_S}{\GInv \wedge \CGPost}}.
                  {\htriple{}{
\begin{array}{c}
\derivedDestroyed{S''} \ast \\
(\derivedCreated{S''} \wand \\
(\CInv' \wedge \CLC' \\
\wedge e_O \wedge e_S))
\end{array}
                  }{\ccalltl{x}{e_r}{fun}{e_a}{e_v}}{
\begin{array}{c}
\aperformed{\derivedPerformed{S''}} \ast \\
\aperformed{S'} \ast
(\oldl{call}{e_O} \wedge \\
\CInv[\oldl{call}{\_} / \old{\_}] \wedge \\
\CGPost[\oldl{call}{\_} / \old{\_}] \wedge \\
Q[e_r/\self][x/ \result ] \\
~[\self/\msg.sender][e_a/x] \\
~[\oldl{call}{\_} / \old{\_}]
\wedge \\
e_N \Rightarrow \oldl{last}{e_N})
\end{array}
                  }} \\
\ \\
\Inf[Seq]{\htriple{}{P}{c_1}{R}}
         {\htriple{}{R}{c_2}{Q}}
         {\htriple{}{P}{\cseq{c_1}{c_2}}{Q}}
\quad \quad
\Inf[Cons]{\htriple{}{P'}{c}{Q'}}{P \entails P'}{Q' \entails Q}
          {\htriple{}{P}{c}{Q}} \\
\ \\
\Inf[Frame]{\FV(R) \cap \mods(c) = \emptyset}{\htriple{}{P}{c}{Q}}.
           {R\text{ is stateless if $c$ contains a call}}
           {\htriple{}{P \ast R}{c}{Q \ast R}} \\
\ \\
\Inf[Func]{\htriple{}{
\begin{array}{c}
\exists l \ldotp \CInv[\oldl{l}{\_}/\old{\_}] \wedge \\
e_{N_1} \Rightarrow \oldl{last}{e_{N_1}} \wedge \\
e_{N_2} \Rightarrow \oldl{pre}{e_{N_2}}
\end{array}
}{c}{
\begin{array}{c}
Q[\oldl{pre}{\_}/\old{\_}] \wedge \\
\CGPost[\oldl{pre}{\_}/\old{\_}] \\
\wedge \CInv[\oldl{last}{\_}/ \old{\_}] \\
\wedge \CLC[\oldl{last}{\_}/ \old{\_}]\\
\ast \aperformed{S}
\end{array}
}}
          {\code{def}~f(x): T~\{c\}~\ensures{Q}~\performs{S}} \\
\ \\
\Inf[Init]{\htriple{}{
          \rcreators \ast \sdefault{\ContractState{\self}}
          }{c}{
\begin{array}{c}
\CInv[\_/\old{\_}] \wedge
Q  \\
\ast \aperformed{S}
\end{array}
          }}.
          {\CInv \text{ precisely determines resource state.}}.
          {\CInv , \CGPost \text{ are reflexive and transitive.}}.
          {c\text{ does not contain any calls.}}.
          {\stable{\CInv}{\GInv \wedge \CCallerPrivate{\self}{\emptyset} \wedge \ContractState{\self} = \old{\ContractState{\self}}}}
          {\code{def}~\code{init}(x)~\{c\}~\ensures{Q}~\performs{S}} \\
\end{array}
\]
\caption{Statement, function and constructor proof rules.}\label{fig:logic1}
\end{figure}

\begin{figure}[ht!]
\[\small
\begin{array}{c}
\Inf[Create]{R \neq \wei}{R\text{ is not derived}}
            {\htriple{}{
\begin{array}{c}
(e_a \neq 0 \Rightarrow \\
\atrusts{e_c}{\code{msg.sender}}{\tru}) \ast \\
\aalloc{\rcreator{R}}{e_c}{1} \\
\wedge e_a \geq 0
\end{array}
}{\ccreate{R}{e_c}{e_t}{e_a}}{
\begin{array}{c}
(e_a \neq 0 \Rightarrow \\
\atrusts{e_c}{\code{msg.sender}}{\tru}) \ast \\
\aalloc{R}{e_t}{e_a} \ast \\
\aperformed{\ecreate{R}{e_t}{e_a}} \ast  \\
\aalloc{\rcreator{R}}{e_c}{1}
\end{array}
}} \\
\ \\
\Inf[Destroy]{R \neq \wei}{R\text{ is not derived}}
            {\htriple{}{
\begin{array}{c}
(e_a \neq 0 \Rightarrow \\
\atrusts{e_f}{\code{msg.sender}}{\tru}) \ast \\
\aalloc{R}{e_f}{e_a}
\wedge e_a \geq 0
\end{array}
}{\cdestroy{R}{e_f}{e_a}}{
\begin{array}{c}
(e_a \neq 0 \Rightarrow \\
\atrusts{e_f}{\code{msg.sender}}{\tru}) \ast \\
\aperformed{\edestroy{R}{e_f}{e_a}}
\end{array}
}} \\
\ \\
\Inf[Transfer]{\htriple{}{
\begin{array}{c}
e_a \geq 0 \ast \\
(a \neq 0 \Rightarrow \\
 \atrusts{e_f}{\code{msg.sender}}{\tru}) \ast \\
\aalloc{R}{e_f}{e_a} \wedge e_a \geq 0
\end{array}
}{\creallocate{R}{e_f}{e_t}{e_a}}{
\begin{array}{c}
\aalloc{R}{e_t}{e_a} \ast \\
(e_a \neq 0 \Rightarrow \\
 \atrusts{e_f}{\code{msg.sender}}{\tru}) \\
\ast \aperformed{\ereallocate{R}{f}{t}{a}}
\end{array}
}} \\
\ \\
\Inf[Trust]{\htriple{}{
\begin{array}{c}
\atrusts{\code{msg.sender}}{e_c}{\_}
\end{array}
}{\ctrust{e_c}{e_v}}{
\begin{array}{c}
\atrusts{\code{msg.sender}}{e_c}{e_v} \\
 \ast \aperformed{\etrust{\code{msg.sender}}{e_c}{e_v}{\self}}
\end{array}
}} \\
\ \\
\Inf[Offer]{R_1 \neq \wei}{\text{No resource derived from $R_1$.}}
           {\htriple{}{
\begin{array}{c}
e_{a_1} \geq 0 \wedge e_{a_2} \geq 0 \\
\wedge e_n \geq 0 \wedge \aemp \ast\\
(e_n \neq 0 \Rightarrow \\
 \atrusts{e_f}{\code{msg.sender}}{\tru})
\end{array}
}{\coffertl{R_1}{R_2}{e_f}{e_t}{e_{a_1}}{e_{a_2}}{e_n}}{
\begin{array}{c}
\aoffers{R_1}{R_2}{e_f}{e_t}{e_{a_1}}{e_{a_2}}{e_n} \\
\ast \code{perf}(\code{offer}_{R_1 \leftrightarrow R_2} \\
(e_f, e_t, e_{a_1}, e_{a_2}, e_n)) \\
\ast (e_a \neq 0 \Rightarrow \\
 \atrusts{e_f}{\code{msg.sender}}{\tru})
\end{array}
}} \\
\ \\
\Inf[Revoke]{\htriple{}{
\begin{array}{c}
e_{a_1} \geq 0 \wedge e_{a_2} \geq 0 \\
\wedge e_n \geq 0 \wedge \aemp \ast\\
(e_n \neq 0 \Rightarrow \\
 \atrusts{e_f}{\code{msg.sender}}{\tru}) \\
 \ast \aoffers{R_1}{R_2}{e_f}{e_t}{e_{a_1}}{e_{a_2}}{e_n}
\end{array}
}{\crevoketl{R_1}{R_2}{e_f}{e_t}{e_{a_1}}{e_{a_2}}{e_n}}{
\begin{array}{c}
(e_a \neq 0 \Rightarrow \\
 \atrusts{e_f}{\code{msg.sender}}{\tru}) \ast \\
 \code{perf}(\code{revoke}_{R_1 \leftrightarrow R_2} \\
 (e_f, e_t, e_{a_1}, e_{a_2}, e_n))
\end{array}
}} \\
\ \\
\Inf[Exchange]{\htriple{}{
\begin{array}{c}
e_{a_1} \geq 0 \ast e_{a_2} \geq 0\\
\ast (e_{a_1} > 0 \Rightarrow \\
\aoffers{R_1}{R_2}{e_f}{e_t}{e_{a_1}}{e_{a_2}}{1}) \\
\ast (e_{a_2} > 0 \Rightarrow \\
\aoffers{R_2}{R_1}{e_t}{e_f}{e_{a_2}}{e_{a_1}}{1}) \\
\ast \aalloc{R_1}{e_f}{e_{a_1}} \ast \aalloc{R_2}{e_t}{e_{a_2}}
\end{array}
}{\cexchangetl{R_1}{R_2}{e_f}{e_t}{e_{a_1}}{e_{a_2}}}{
\begin{array}{c}
\aalloc{R_1}{e_t}{e_{a_1}} \ast \\
\aalloc{R_2}{e_f}{e_{a_2}} \ast \\
\code{perf}(\code{exchange}_{R_1 \leftrightarrow R_2} \\
 (e_f, e_t, e_{a_1}, e_{a_2}, 1))
\end{array}
}}
\end{array}
\]
\caption{Rules for resource ghost commands. }\label{fig:logic2}
\end{figure}

The rules for ordinary statements are standard; the bulk of the work happens in the rule for calls, as well as in the rules for resource commands. We will explain the checks for every element of our methodology step by step.

Ordinary and transitive segment constraints must hold at the end of every local segment. We therefore require them to hold at the end of each function as well as before every call (in a state where the Ether that is about to be sent with the call has already been deducted), \ie{} in the premise of the call rule (we will discuss the resource assertions here later), \wrt{} the old state at the beginning of the local segment (\ie{} the old state labelled ``last''). Crucially, what this last state is does not stay constant throughout a function, but it changes every time a new local segment starts, \ie{} after each call. To model the fact that after a call, the new ``last'' state is the current state, the postcondition of the call rule allows assuming that any expression $e_N$ that is true in the current state is also true in the last state. Similarly, the rule for functions allows doing this at the beginning of each function.

The frame rule ensures that no information about the last state or the current state can be framed around calls; this represents the fact that the entire contract state can change with every call.
After a call, one may assume the transitive segment constraints \wrt{} to the call's pre-state. To remember information about said pre-state, we use the same trick as before, and allow assuming any expression $e_O$ after a call about its pre-state that was known to be true before the call.
In constructors, no calls are allowed, and we check at the end that transitive segment constraints hold in the current state \wrt{} itself, which ensures that all single-state invariants contained in the transitive segment constraints are established.

The proof obligations for function constraints work in a similar way: They must be shown to hold at the end of each function \wrt{} to its pre-state (which may again be remembered by assuming that everything that holds in the beginning of the function holds in its pre-state), and may be assumed after a call \wrt{} to the call's pre-state.

The rule for constructors ensures that all transitive segment constraints and function constraints are, in fact, reflexive and transitive, and that the resource state is precisely determined by the transitive segment constraints (\ie{} they precisely determine the resource state in terms of the contract state).
Similarly, the rule for the constructor ensures that transitive segment constraints are stable under the transitive segment constraints and privacy constraints of all known interfaces. For every call to a secondary contract, the call rule requires that the transitive segment constraints, potentially \emph{strengthened} by conjoining them with some arbitrary information $e_S$ that is known about the current state (like the fact that \code{self.lock} is set in the example in Sec.~\ref{sec:ici}), are stable under each of the following two assertions:
\begin{enumerate}
  \item The conjunction of the transitive segment constraints of all interfaces, the knowledge that the primary contract's state does not change, and the privacy constraints of all secondary contracts \emph{except} the called one (which is now trivial).
  \item The conjunction of the transitive segment constraints of all interfaces and the function constraints of the primary contract.
\end{enumerate}
Fig.~\ref{fig:secondary-call} illustrates the scenario where the primary contract calls a secondary contract. The strengthened invariant is known to hold in state 1. Whatever local changes the secondary contract performs cannot break the strengthened invariant, since it is stable under the first assertion (\eg{} in the example, changes in the secondary contract cannot break the strengthened invariant that remains true as long as \code{self.lock} is set).
Any contracts executing between states 2 and 5 cannot break the original non-strengthened invariant by the reasoning laid out in Sec.~\ref{sec:ici}.
Additionally, because the strengthened invariant is stable under the second assertion, it will also be re-established by the time any re-entrant calls the secondary contract may (transitively) make on the primary contract return (states 3 and 4); in the example, a function constraint guarantees that \code{self.lock} will again be set when such a re-entrant call returns. As a result, again because of stability under the first assertion, the secondary contract again cannot break the invariant after the return (between states 5 and 6) either.

\begin{figure}
\includegraphics[width=0.8\textwidth]{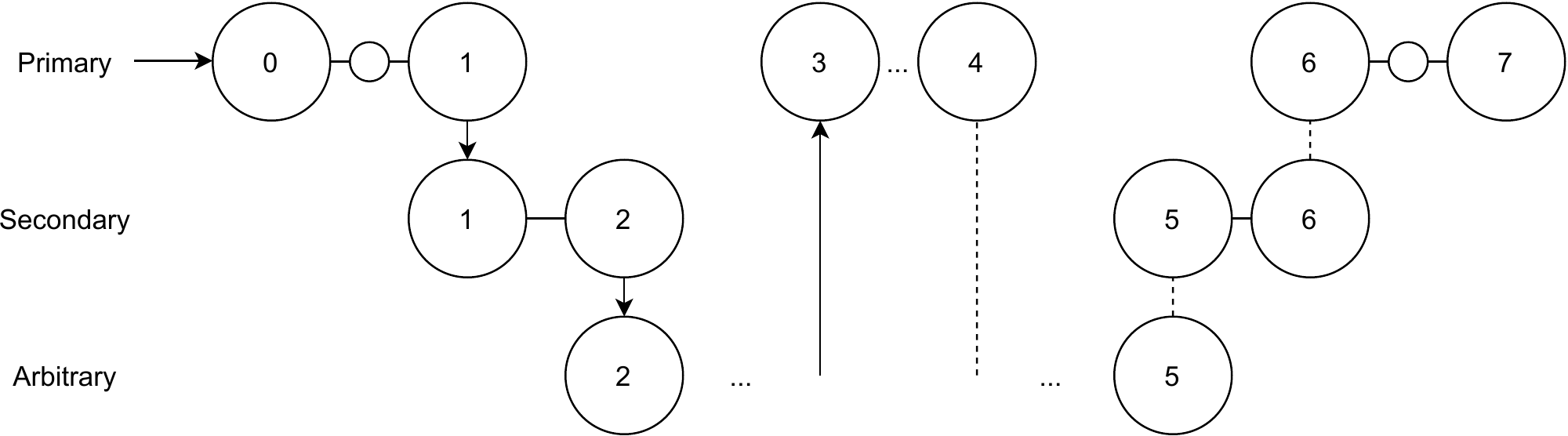}
\caption{Example showing a call from a primary to a secondary contract that leads to a re-entrant call to the primary contract.}\label{fig:secondary-call}
\end{figure}

The rules for resource commands are mostly intuitive. A transfer, for example, requires that the sender initially has the resources that are to be transferred, and ends in a state where the recipient has them. It also requires that the caller of the function is trusted by the address whose resources are being transferred (everybody always trusts themselves). Finally, the conclusion states that a transfer-effect has occurred. Rules for other resource commands are analogous; for example, the rule for exchanges requires two compatible offers (unless the offered amount of a party is zero) and consumes them, switches ownership of the involved resources, and records that an exchange-effect has occurred.
The rule for resource creation requires that $e_c$ (the party who is creating the new resource of type $R$) owns a creator-resource for $R$, which represents the right to create new resources.
These creator resources can be created and given to arbitrary addresses in the contract's constructor:
In the constructor rule, the caller of the constructor is given the right to create such creator-resources for any resource the contract declares. This is denoted by $\rcreators$, which, for a contract with resources $R_0, \dots, R_n$, is defined as $\aalloc{\rcreator{\rcreator{R_0}}}{\code{msg.sender}}{1} \ast \dots \ast \aalloc{\rcreator{\rcreator{R_n}}}{\code{msg.sender}}{1}$.

The function and constructor rules ensure that, at the end of the function or constructor, the multiset of recorded effects is exactly that which has been declared.

Finally, calls can also interact with the resource state. First, any subset of the effects declared by the called function may be recorded by the caller. Second, if a called function declares resource effects w.r.t. a resource $R$ s.t. the current contract has a resource $D$ derived from $R$, then these effects lead to implicit effects on the derived resources. For example, transferring $R$ away to someone implicitly destroys the same number of $D$ they must currently own, and requires that they have offered to exchange that amount of $D$ for the same amount of $R$. This is captured by the functions $\derivedCreated{}$ and $\derivedDestroyed{}$, the former of which defines which derived resources are implicitly created by an effect, and the latter defines which derived resources are implicitly destroyed. Their definitions can be found in Fig.~\ref{fig:derived-defs}. The premise of the call rule ensures that transitive segment constraints etc. hold in a state where destroyed derived resources have already been removed and created derived resources have already been added.
The $\derivedDestroyed{}$ clause also ensures that no offers are made to give away resource $R$ by a contract that defines $D$, and that the contract cannot trust someone \wrt{} to the contract that declares $R$; either of these could lead to some amount of $R$ being removed from the contract without the appropriate checks that the receiver has sufficient amounts of the derived resource $D$.
Finally, the call-rule ensures that all such implicit effects on derived resources (defined by $\derivedPerformed{}$) are also recorded.

\begin{figure}
\[\small
\begin{array}{rcl}
\derivedCreated{\ereallocate{R}{f}{\self}{a}} &=& \aalloc{D}{f}{a} \\
\derivedCreated{\eexchange{R}{R'}{f}{\self}{a_1}{a_2}{n}} &=& \aalloc{D}{f}{n*a_1}  \\
\derivedCreated{\ecreate{R}{\self}{a}} &=& \aalloc{D}{\code{msg.sender}}{a} \\
\derivedCreated{\_} &=& \aemp \\
\derivedDestroyed{\ereallocate{R}{\self}{t}{a}} &=& \aalloc{D}{t}{a} \ast \aoffers{D}{R}{t}{\self}{1}{1}{a} \\
\derivedDestroyed{\edestroy{R}{\self}{a}} &=& \aalloc{D}{\code{msg.sender}}{a} \ast  \\
& & \aoffers{D}{R}{\code{msg.sender}}{\self}{1}{1}{a} \\
\derivedDestroyed{\eoffer{R}{R'}{\self}{t}{a_1}{a_2}{n}} &=& \fal \\
\derivedDestroyed{\etrust{\self}{t}{v}{c}} &=& \fal \text{ if $c$ declares $R$} \\
\derivedDestroyed{\_} &=& \aemp \\
\derivedPerformed{\ereallocate{R}{f}{\self}{a}} &=& \multiset{\ecreate{D}{f}{a}} \\
\derivedPerformed{\eexchange{R}{R'}{f}{\self}{a_1}{a_2}{n}} &=& \multiset{ \ecreate{D}{f}{n*a_1} }\\
\derivedPerformed{\ecreate{R}{\self}{a}} &=& \multiset{ \ecreate{D}{\code{msg.sender}}{a} } \\
\derivedPerformed{\ereallocate{R}{\self}{t}{a}} &=& \multiset{ \edestroy{D}{t}{a} }\\
\derivedPerformed{\edestroy{R}{\self}{a}} &=& \multiset{ \edestroy{D}{\code{msg.sender}}{a} }\\
\derivedPerformed{\_} &=& \emptymultiset
\end{array}
\]
\caption{Functions describing the implicit consequences of the effects of called functions on derived resources. $D$ is assumed a resource derived from $R$. We write $\multiset{...}$ for multiset literals.} \label{fig:derived-defs}
\end{figure}

\section{Extended Examples}

Here, we show (extended versions of) the auction and token examples, annotated with different kinds of specifications. The syntax is similar to that of \twovyper, but simplified for presentation reasons. For the same reason, we have omitted some functions, some parts of the code (in particular, the locks) and some parts of the specification.

\subsection{Auction example}\label{app:example_auction}
The following is a slightly simplified version of the auction contract annotated with resource specifications that use a derived resource. Note that in our tool, a derived resource for wei is declared automatically (meaning that by default, the assumption is that wei sent to a contract should still conceptually belong to the sender, \ie{} they get a title for it); here, we have explicitly declared it with the name \code{wei_in_auction} for illustration purposes.

\begin{lstlisting}[numbers=none]
beneficiary: address
highestBid: int256
highestBidder: address
ended: bool
pendingReturns: map(address, int256)

#@ resource: good()
#@ resource: wei_in_auction() derived from wei

#@ transitive segment constraint: ... # relate contract state and resource state

#@ segment constraint: msg.sender != self.beneficiary ==> self.ended == old(self.ended)
#@ transitive segment constraint: self.beneficiary == old(self.beneficiary)
#@ transitive segment constraint: old(self.ended) ==> self.ended

#@ performs: create[wei_in_auction](msg.value)
#@ performs: offer[wei_in_auction <-> good](msg.value, 1, to=self.beneficiary, times=1)
@public
@payable
def bid():
    assert block.timestamp < self.auctionEnd and not self.ended
    assert msg.value > self.highestBid and msg.sender != self.beneficiary

    #@ offer[wei_in_auction <-> good](msg.value, 1, to=self.beneficiary, times=1)

    self.pendingReturns[self.highestBidder] += self.highestBid
    self.highestBidder = msg.sender
    self.highestBid = msg.value


#@ performs: destroy[wei_in_auction](self.pendingReturns[msg.sender])
@public
def withdraw():
    pending_amount: wei_value = self.pendingReturns[msg.sender]
    self.pendingReturns[msg.sender] = 0
    send(msg.sender, pending_amount)


#@ performs: exchange[wei_in_auction <-> good](self.highestBid, 1, self.highestBidder,
#@                                             self.beneficiary, times=1)
#@ performs: destroy[wei_in_auction](self.highestBid, actor=self.beneficiary)
@public
def endAuction():
    assert block.timestamp >= self.auctionEnd and not self.ended
    self.ended = True

    #@ exchange[wei_in_auction <-> good](self.highestBid, 1, self.highestBidder,
    #@                                   self.beneficiary, times=1)

    send(self.beneficiary, self.highestBid)
\end{lstlisting}

\subsection{Token example}\label{app:example_token}
The following is an extended version of the token contract (namely one that is close to a real implementation of ERC20 in Vyper) using resource specifications and effects-clauses. In our tool, and therefore in this example, effects-clauses are introduced with the keyword \code{performs}. The code also contains segment constraints that constrain when \emph{events} are triggered; for a brief description, see our evaluation in the main body of the paper.

\begin{lstlisting}[numbers=none]
minter: address
balances: map(address, int256)
allowances: map(address, map(address, int256))

#@ resource: token()

#@ transitive segment constraint: self.minter == old(self.minter)
#@ transitive segment constraint: self.total_supply == sum(self.balances)

#@ transitive segment constraint: allocated[token]() == balanceOf(self)
#@ transitive segment constraint: forall({o: address, s: address},
#@   self.allowances[o][s] == offered[token <-> token](1, 0, o, s))

#@ segment constraint: forall({a: address, b: address},
#@   self.balanceOf[a] > old(self.balanceOf[a]) and self.balanceOf[b] < old(self.balanceOf[b])
#@     ==> event(Transfer(b, a, self.balanceOf[a] - old(self.balanceOf[a]))))
#@ ... (omitted, more specifications for events)


#@ performs: create[token](_value, to=_to)
@public
def mint(_to: address, _value: uint256):
    assert msg.sender == self.minter
    assert _to != ZERO_ADDRESS
    self.total_supply += _value
    #@ create[token](_value, to=_to)
    self.balanceOf[_to] += _value
    log.Transfer(ZERO_ADDRESS, _to, _value)

#@ performs: transfer[token](_value, to=_to)
@public
def transfer(_to: address, _value: uint256) -> bool:
    self.balanceOf[msg.sender] -= _value
    #@ transfer[token](_value, to=_to)
    self.balanceOf[_to] += _value
    log.Transfer(msg.sender, _to, _value)
    return True

#@ performs: exchange[token <-> token](1, 0, _from, msg.sender, times=_value)
#@ performs: transfer[token](_value, to=_to)
@public
def transferFrom(_from: address, _to: address, _value: uint256) -> bool:
    self.balanceOf[_from] -= _value
    self.balanceOf[_to] += _value
    self.allowances[_from][msg.sender] -= _value
    #@ exchange[token <-> token](1, 0, _from, msg.sender, times=_value)
    #@ transfer[token](_value, to=_to)
    log.Transfer(_from, _to, _value)
    return True

#@ performs: revoke[token <-> token](1, 0, to=_spender)
#@ performs: offer[token <-> token](1, 0, to=_spender, times=_value)
@public
def approve(_spender: address, _value: uint256) -> bool:
    #@ revoke[token <-> token](1, 0, msg.sender, _spender,
    #@                         offered[token <-> token](1, 0, msg.sender, spender))
    self.allowances[msg.sender][_spender] = _value
    #@ offer[token <-> token](1, 0, to=_spender, times=_value)
    log.Approval(msg.sender, _spender, _value)
    return True
\end{lstlisting}

\section{Further Verified Examples}\label{app:other-examples}

Here, we describe the properties we proved for the contracts not explicitly mentioned in the main body of the paper.

\myparagraph{ERC721:}
ERC721 is a more complex token standard than ERC20. We declared that it implements a token resource whose tokens have identifiers (non-fungible tokens, NFTs), and specified its functions in terms of token transfers and exchanges. Like ERC20, this required using offers and exchanges, but in addition, it also required using trust, since ERC721 allows users to name other users as ``operators'' who can act on their behalf.

\myparagraph{Serenuscoin:}
Here we use segment constraints prove both access control properties (only the owner may change the factory) and that the correct events are triggered under the right circumstances.

\myparagraph{Mana:}
We verified a simplified version of the Mana application from the VerX paper, where we focused on the parts necessary to show inter-contract invariants between the three collaborating contracts that were also verified (non-modularly) by VerX. Some of these are not single-state invariants but two-state inter-contract transitive segment constraints; one example is that once the token contract's owner has been set to be the \code{continuous_sale} contract, its owner will never change again.

\myparagraph{Safe remote purchase:}
This smart contract sells a good to an arbitrary buyer and holds the buyer's funds in escrow until they acknowledge that they have received the good. The contract gives both parties an incentive not to block the other party from receiving funds by holding a deposit from each of them.
We use a derived resource for wei (which, as we stated above, our tool declares by default) to model the fact both buyer and seller conceptually own their deposits until the transaction is finalized, at which point the buyer's wei-titles are exchanged for the good, and the deposits can be paid back.

\fi

\end{document}